\documentclass[final, a4paper]{manuscript} 
\usepackage{amscd,braket}
\usepackage{amsfonts,dsfont}
\usepackage{tikz,subfigure}
\usepackage{soul,cancel}
\usepackage{comment}
\numberwithin{equation}{section}

\def\p{\partial}
\newcommand{\bea}{\begin{eqnarray}}
\newcommand{\eea}{\end{eqnarray}}
\newcommand{\be}{\begin{equation}}
\newcommand{\ee}{\end{equation}}
\newcommand{\ba}{\begin{align}}
\newcommand{\ea}{\end{align}}
\newcommand{\zl}{ z^{\textrm{L}}_{\textrm{min}}}

\newcommand{\Tr}{{\rm {Tr}}}

\newcommand\rref[1]{(\ref{#1})}

\newcommand{\ie}{{\it i.e.~}}

\newlength{\slength}
\manunotes{PN}
\manunotes{JS}
\manunotes{AB}
\manunotes{JdB}
\manunotes{DJa}

\title{Approximate CFTs and Random Tensor Models}
\author[a,b,c,d]{Alexandre Belin}
\author[e]{Jan de Boer}
\author[f]{Daniel L. Jafferis}
\author[a,g]{Pranjal Nayak}
\author[a]{Julian Sonner}

\affiliation[a]{D\'epartment de Physique Th\'eorique, Universit\'{e} de Gen\`eve, \\24 quai Ernest-Ansermet, 1211 Gen\`eve 4, Suisse}
\affiliation[b]{Dipartimento di Fisica, Universit\`a di Milano - Bicocca \\
I-20126 Milano, Italy}
\affiliation[c]{INFN, Sezione di Milano-Bicocca, Piazza della Scienza 3, 20126 Milano, Italy}
\affiliation[d]{Institute of Physics, Ecole Polytechnique F\'ed\'erale de Lausanne, \\ CH-1015 Lausanne, Switzerland}
\affiliation[e]{Institute for Theoretical Physics, University of Amsterdam, \\PO Box 94485, 1090 GL Amsterdam, The Netherlands}
\affiliation[f]{Department of Physics, Harvard University, Cambridge, MA 02138, USA}
\affiliation[g]{Department of Theoretical Physics, CERN, \\Esplanade des Particules 1, 1211 Gen\`eve 23, Suisse}

\emailAdd{alexandre.belin@unimib.it}
\emailAdd{J.deBoer@uva.nl}
\emailAdd{jafferis@g.harvard.edu}
\emailAdd{pranjal.nayak@cern.ch}
\emailAdd{julian.sonner@unige.ch}

\preprint{CERN-TH-2023-068}

\abstract{

A key issue in both the field of quantum chaos and quantum gravity is an effective description of chaotic conformal field theories (CFTs), that is CFTs that have a quantum ergodic limit. We develop a framework incorporating the constraints of conformal symmetry and locality, allowing the definition of ensembles of `CFT data'. These ensembles take on the same role as the ensembles of random Hamiltonians in more conventional quantum ergodic phases of many-body quantum systems. To describe individual members of the ensembles, we introduce the notion of approximate CFT, defined as a collection of `CFT data' satisfying the usual CFT constraints approximately, i.e. up to small deviations. We show that they generically exist by providing concrete examples. Ensembles of approximate CFTs are very natural in holography, as every member of the ensemble is indistinguishable from a true CFT for low-energy probes that only have access to information from semi-classical gravity. To specify these ensembles, we impose successively higher moments of the CFT constraints. Lastly, we  propose a theory of pure gravity in AdS$_3$ as a random matrix/tensor model implementing approximate CFT constraints. This tensor model is the maximum ignorance ensemble compatible with conformal symmetry, crossing invariance, and a primary gap to the black-hole threshold. The resulting theory is a random matrix/tensor model governed by the Virasoro 6j-symbol.

}

\begin{document}
\maketitle

\section{Introduction}\label{sec.intro}
The study of low-dimensional models of gravity has led to important insights for some of the biggest puzzles in quantum gravity, notably for the black hole information paradox \cite{Penington:2019npb,Almheiri:2019psf,Penington:2019kki,Almheiri:2019qdq}. Most of the progress has come from studying simple and UV complete gravitational theories in (nearly) AdS$_2$, which have a striking feature that distinguish these from higher dimensional examples of AdS/CFT. In two dimensions, holographic duality does not  seem to involve a single boundary quantum system with a fixed Hamiltonian, but rather an ensemble average over Hamiltonians, that is a matrix model where the random matrix has the interpretation of the boundary Hamiltonian. A precursor for this observation was the relation between the IR sector of the SYK model \cite{Kitaev-talks:2015, Maldacena:2016hyu, Maldacena:2016upp} and AdS$_2$. A precise version of such a duality is that between pure JT gravity in two dimensions and a specific matrix model \cite{Saad:2019lba}, which is valid even at the (doubly) non-perturbative level \cite{Johnson:2019eik,Post:2022dfi,Altland:2022xqx}. While it remains an interesting question whether low-dimensional holography -- in this case in two dimensions -- can be formulated for a fixed boundary quantum system, it is clearly imperative to understand the role of holographic dualities which involve ensemble averages in dimensions higher than two. Here the situation, at least at first sight, is reversed: known examples involve dualities between individual quantum systems, such as ${\cal N}=4$ SYM in four dimensions being dual to IIB string theory on AdS$_5\times$S$^5$. Moreover we do not presently have a well-developed theory of the appropriate candidate ensembles of higher-dimensional field theories that could possibly be relevant in such cases.\footnote{There are cases where CFTs have conformal manifolds that can be averaged over using the Zamolodchikov metric as a measure \cite{Maloney:2020nni,Afkhami-Jeddi:2020ezh,Collier:2022emf} (see also \cite{Heckman:2021vzx}). However, there are either too few marginal directions (like in $\mathcal{N}=4$ SYM) or the theories are not holographic (like the Narain CFTs). Moreover, there are cases of holographic CFTs where it is clear that there are no marginal directions at all (like the 6d $(2,0)$ theory).} Thus an interesting open question, at least on the field-theory side is: how do we construct ensembles of local quantum field theories (including CFTs) which could take on a role similar to the matrix model dual of JT gravity in two dimensions?\footnote{See \cite{Douglas:2010ic} for earlier attempts on developing a measure on the space of CFTs.} We will address the issue of whether such ensembles should be fundamental, or rather to be understood as emergent in the sense of quantum chaos further below.

On the gravity side much of this progress is driven by a key insight: including Euclidean (or spacetime) wormholes in the gravitational path integral over metrics gives access to novel and non-perturbative information about the dual system, in particular on the statistical distribution of black-hole microstates.\footnote{Moreover, such geometries, in the guise of `replica wormholes' come to the rescue of unitarity at late times \cite{Saad:2019lba,Penington:2019npb,Almheiri:2019psf,Penington:2019kki,Almheiri:2019qdq} and give a semi-classical Page curve in line with unitary expectations.}\footnote{This statistical distribution can also be obtained from one-boundary computations if one combines these with a maximal entropy principle in the spirit of statistical physics. Thus, wormholes can also be seen as a reaffirmation of the validity of the maximal entropy principle \cite{toappear}.} However, conceptually, the inclusion of such multi-boundary connected geometries raise important physical questions, which were already hinted at above, and which are sometimes referred to as the `factorization puzzle', since one of the signatures of their inclusion is a non-factorization of quantities which ought to decompose into disconnected single-boundary quantities, for example the product of two partition functions, $Z(\beta_1,\beta_2) \Big|_{\textrm{grav}} \neq Z_{\textrm{CFT}}(\beta_1) Z_{\textrm{CFT}}(\beta_2)$. A possible interpretation of this phenomenon is that gravity is dual to an ensemble average over quantum systems, also in higher dimensions \cite{Marolf:2020xie}. An alternative possibility is that semi-classical gravity only captures a coarse-grained version of the true microscopic observables, and that this coarse-graining is responsible for the lack of factorization \cite{Pollack:2020gfa,Belin:2020hea,Altland:2020ccq}. This latter point of view is rooted in the framework of quantum chaos, where analogous phenomena are well understood. From this perspective, one expects that most observables will display erratic oscillations at late time, while semi-classical gravity only captures the mean (as well as the higher moments) of the signal.

As is well understood since the seminal works of Wigner \cite{WignerRMT}, modelling a complex quantum system by an ensemble of theories, traditionally a random matrix theory (RMT), offers an efficient description of the system under consideration, but of course makes no claim about the ensemble description being fundamental. In fact, the RMT description applies to the system, even though it has an explicit underlying description in terms of a single (chaotic) Hamiltonian, and the interpretation of the random-matrix distribution in this context is precisely that it captures the various moments in a statistical description of the single underlying quantum system. The question of which RMT description applies to which underlying quantum system has been solved for many-body quantum systems and is summarised in the Altland-Zirnbauer (AZ) classification of RMT universality classes \cite{Verbaarschot, Verbaarschot.Zahed, Altland:1997zz}. At first sight, and at the level of anti-unitary symmetries which determine which AZ class a given system is in, this classification continues to apply for higher-dimensional quantum field theories, including the strongly coupled conformal field theories relevant in holographic duality. In fact, one expects that the fully ergodic late-time physics of such theories are described block-wise\footnote{Blocks correspond to particular values of a maximal set of commuting conserved charges.} by RMT universality, see e.g. \cite{Cotler:2020hgz,Belin:2020jxr,Belin:2021ibv,Haehl:2023tkr,DiUbaldo:2023qli} for applications of this idea to conformal field theories, and \cite{Delacretaz:2022ojg} for a numerical study of spectral statistics in quantum field theory. However, local quantum field theories, and conformal field theories in particular have much richer structures, which are not captured in such an approach.  We would like to retain these structures from the point of view of a more detailed quantum-chaotic description of such systems, as well as from the point of view of holographic duality. 

In the case of conformal field theories, such locality and symmetry constraints can be efficiently formulated in the form of the so-called CFT or `bootstrap' constraints \cite{Rattazzi:2008pe} (see e.g. \cite{Simmons-Duffin:2016gjk} for a review). The list of CFT data, that is the set of conformal dimensions, spins and structure constants, $\{\Delta_i, J_i, C_{ijk} \}$, do not define a consistent CFT unless these data satisfy a number of `crossing constraints', which will feature prominently in this paper. The most well-known among these expresses the equivalence of the conformal block decomposition of a correlation function of four operators in two different contraction channels, which for simplicity we give for four identical external operators,
\begin{equation}\label{eq.STCrossingFirstMention}
    \sum_{\Delta, {\ell}} p_{\Delta, \ell} \left(u^\Delta g_{\Delta, \ell}(u,v) - v^\Delta g_{\Delta, \ell} (v,u)  \right) = 0\,,
\end{equation}
where the $g_{\Delta, \ell}$ are global conformal blocks, $u$ and $v$ are cross-ratios, and $p_{\Delta,\ell}$ are positive coefficients depending on the structure constants $C_{ijk}$. Only CFT data which exactly satisfy this constraint, and a number of related ones actually define a local, conformally invariant quantum field theory, i.e a CFT. Throughout this paper we will refer to this set of constraints as `CFT constraints', although we will later (need to) be more precise about which set of constraints we are talking about.

The main goal of this paper is to describe a new approach that incorporates the CFT bootstrap into chaos universality, allowing us to give a mathematical description of the quantum ergodic behavior of conformal field theories in ways that go beyond the usual AZ classification and that take all relevant CFT constraints into account. This involves first defining an  `approximate CFT', which will take on the role of an individual element of an ensemble of `approximate CFTs'. This ensemble of approximate CFTs can then be used to construct quantum chaotic ensembles of CFTs which respect the bootstrap constraints in an approximate sense to be specified later. In the latter part of the paper we will then specialise to two-dimensional conformal symmetry, including a careful treatment of the infinite-dimensional Virasoro symmetry, and define the ensemble of two-dimensional chaotic approximate CFTs. Restricting to the operator content believed to be relevant for CFT duals of pure gravity, we will develop a probability distribution of CFT data -- appropriately regulated -- that resembles a discrete description of three-dimensional Euclidean spacetime in terms of random tensors. The tensorial part of this model provides a statistical distribution of CFT structure constants whose non-Gaussianities are controled by the Ponsot-Teschner inversion kernel. We now turn to a slightly more detailed summary of the main findings of this paper. In reading this paper, it is important to keep in mind that we approach the idea of approximate and average CFTs in three different ways, which in order in which they appear in the paper are
\begin{enumerate}\setcounter{enumi}{1}
    \item Through the specification of an individual element of the ensemble, namely by defining an `approximate CFT'.
    \item By specifying the moments of the ensemble of approximate CFTs, which will turn out to be necessarily non-Gaussian, a property implied by studying the variance of the crossing equation.
    \item By constructing an explicit probability distribution over structure constants, and operator dimensions.\footnote{We comment on the issue of spin in the corresponding section, Sec. \ref{sec.tensor}.} For the case of a 2D CFT, this results in a random matrix and tensor model, governed by the 6j-symbol of the Virasoro algebra which connects to the simplicial approach of 3D quantum gravity.
\end{enumerate}
Note that we have chosen the numbers before each item to match the corresponding sections of the paper for ease of comparison and we use the same numbering convention for the three subsections that follow, each giving an introduction to the corresponding section of the paper.

\setcounter{subsection}{1}
\subsection{Approximate CFTs and Averaging}

We specify an approximate CFT by the same information as an exact CFT, namely by a list of local operators with conformal dimension and spin $\{\Delta_i,J_i\}$, as well as OPE coefficients $C_{ijk}$. Unlike the case of an exact CFT, this data satisfies the bootstrap constraints only approximately, and only for a restricted set of observables. In particular, the amount of approximation is quantified in terms of the following parameters:
\be
\mathcal{C}_{\textrm{approx}}\rightarrow \{n_{\textrm{max}}, \Delta_{\textrm{max}},  g_{\textrm{max}}, z^{\textrm{L}}_{\textrm{min}} ,\mathbb{T}\} \,,
\ee
where $n_{\textrm{max}}$ puts an upper bound on the number of local operators in a correlation function. $\Delta_{\textrm{max}}$ puts an upper bound on the scaling dimension of the external operators whose correlation function reliably obey the CFT constraints up to exponentially small corrections, the size of which is given by the parameter, $\mathbb T$. Here, $z^{\textrm{L}}_{\textrm{min}}$ is a cut-off in Lorentzian kinematics.\footnote{For correlation function with more than four operators, there would be multiple Lorentzian cross-ratios and each of them would be restricted. We will refer to $z^{\textrm{L}}_{\textrm{min}}$ as a bound on all these cross-ratios.} We show in section \ref{sec.example-approxCFT} that no cutoff is necessary in purely Euclidean kinematics. 
Finally, for CFTs in two dimensions, $g_{\textrm{max}}$ puts an upper bound on the genus of Riemann surfaces on which we compute correlation functions, and $z^{\rm L}_{\rm min}$ should be generalized to imply restrictions on the moduli of the Riemann surfaces.

For the case of 2d CFTs, modular invariance on higher genus surfaces has a nice interpretation in terms of four-point crossing of operators above $\Delta_{\max}$. For example, modular invariance at genus-two is a crossing of four heavy operators, further summed over all external heavy operators. Imposing approximate modular crossing at genus-two implies that crossing for almost all heavy operators has to be obeyed, up to the tolerance\footnote{More precisely, a smeared version of heavy operator crossing symmetry has to be valid up to the tolerance. Here the ``smearing'' is similar to the smearing with $e^{-\beta E}$ of the spectral density to produce the partition function. A more quantitative statement would require an investigation of the way errors behave under Laplace transforms and generalizations thereof.}. It is trickier to do this in higher dimensions, because the partition function on general manifolds cannot be computed in terms of the local CFT data (even in principle). One would need to find the right construction (see \cite{Benjamin:2023qsc} for work in this direction) in order to derive approximate four-point crossing for almost all heavy operators.

\subsubsection*{Motivation from semi-classical gravity}
Much of the motivation for considering such a framework comes from gravity. In AdS/CFT, semi-classical gravity represents the low-energy effective theory of the bulk description. It is well-tailored to compute low-point correlation functions in the vacuum or in thermal states, but there are quantities that it cannot probe. For example, it cannot be used to compute an $N$-point correlation function in the vacuum, where `$N$' is a generic label for the number of local degrees of freedom, since with that number of external operators, gravity becomes strongly coupled.\footnote{See however \cite{Anous:2016kss,Anous:2017tza} for a calculation of an arbitrary point function. Note, however, that the individual weights of the operators are scaled in a particular way, such that $\Delta\sim \epsilon N$, with $\epsilon \to 0$ and $N\to\infty$.} Similarly, it cannot compute four-point functions of black hole microstates (here we mean energy eigenstates). In fact, one cannot even set-up that experiment within semi-classical gravity since there is no way of specifying which black hole microstate we are trying to scatter. However, when considering Euclidean theories in two dimensions, non-trivial information about such scattering between external black-hole (micro-)states can be obtained by using modular invariance of the dual CFT defined on higher genus surfaces, and correspondingly Hawking-Page like phase transitions in the gravitational theory. 
Building on the holographic intuition, a natural choice for the parameters of an approximate CFT is 
\be
n_{\textrm{max}}\sim N \,, \qquad \Delta_{\textrm{max}} \sim N \,.
\ee
Similarly, low-point correlation functions can be effectively strongly coupled in Lorentzian kinematics, when they produce a collision with center of mass energy that reaches the Planck scale. The time-scale for such a process is the scrambling time \cite{Maldacena:2015waa,Roberts:2014ifa}, so we should introduce a cutoff in Lorentzian kinematics given by
\be
\zl \sim \log N\,.
\ee
For two-dimensional CFTs, we would also expect a break-down of bulk effective field theory if the genus of the Riemann surface on which the CFT lives becomes too large, but the precise scaling is unknown. A natural possibility is that
\be
g_{\max}\sim N \,.
\ee

\subsubsection*{Island-averaging and non-factorization}
As is well-known from the conformal bootstrap program, imposing crossing contraints on the CFT data leads to restricted regions of the parameter space, sometimes referred to as ``islands''.
Similarly for us, once a set of choices of the approximate CFT parameters has been made, we expect to find an ``island'' of parameter space that satisfies the set of constraints we have imposed.\footnote{Calling this region an island is an abuse of terminology, since the topology of the infinite parameter space we are considering may be extremely complicated.} We can then perform an ensemble average over the island with a particular choice of measure, $P$, so that the ensemble average of an observable $\mathbb{O}$ gives
\be\label{eq.avgobs}
\overline{\mathbb{O}}= \int\limits_{\textrm{island}} \!\!d\Delta_i dC_{ijk} P(\Delta_i, C_{ijk}) \mathbb{O}(\Delta_i,C_{ijk}) \,.
\ee
The measure that defines the ensemble above is fixed by imposing various constraints of the physical system under consideration. In the simpler context of JT gravity with matter, this idea was used to derive the measure in \cite{Jafferis:2022uhu,Jafferis:2022wez}. In the case of 2d CFTs imposing CFT constraints like modular invariance on higher genus surfaces, as well as requiring crossing invariance of the correlation functions restricts the measure of the ensemble. One of the goals of this work consists in studying and deriving a measure that satisfies the CFT constraints to a specified level of accuracy, which will necessarily lead to non-linearities, that is to a non-Gaussian statistical distribution for the dynamical data.

The ensemble is labelled by the choice of the parameters defining the set of approximate CFTs, along with the probability measure over the island. Here, we hope to lay down the foundations of approximate CFTs and their associated ensembles and describe certain universal features that are relevant to understand quantum chaos. In the context of holographic theories, we make important starting assumptions: the ensemble should have a large central charge, a large gap to higher spin operators \cite{Heemskerk:2009pn,Afkhami-Jeddi:2016ntf,Meltzer:2017rtf,Belin:2019mnx,Kologlu:2019bco}, and a sparse number of light operators. These assumptions can extend the regime of validity of universal CFT features. Following \cite{Hartman:2014oaa}, these statements can be made precise in two dimensions where the universal formulas can extend all the way down to the states at the black hole threshold.

It is clear that in ensembles of approximate CFTs, products of observables do not factorize
\be\label{eq.nonFactorization}
\overline{\mathbb{O}_1 \mathbb{O}_2}\neq \overline{\mathbb{O}_1} ~ \overline{\mathbb{O}_2} \,.
\ee
In gravity, the non-factorizing property of the ensemble \eqref{eq.nonFactorization} has a geometric interpretation as connected multi-boundary spacetimes \cite{Saad:2019lba,Altland:2020ccq}. Indeed, by appropriately tuning the ensemble one can match wormhole calculations performed in semi-classical gravity \cite{Belin:2020hea,Cotler:2020ugk,Chandra:2022bqq}.

As we have mentioned previously, the notion of approximate CFTs will be useful to characterize quantum chaotic CFTs, independently of holographic applications. What is the right ensemble to draw from if we wish to characterize CFTs in the regime of random matrix universality? As we have emphasised above, it cannot be a simple ensemble of Hamiltonians as is done in quantum mechanics, because of the many bootstrap constraints. However, we propose that only a small number of constraints are enough to characterize the universality class, and as such we can sample over all CFT data that satisfies this subset of constraints.

\subsection{The variance of the crossing equation}
Let us now describe some of our results regarding the particular  constraint of the crossing equation mentioned at the beginning of this introduction, Eq. \eqref{eq.STCrossingFirstMention}.
Crossing invariance in this sense follows from the associativity of the OPE expansion within a correlation function, which means that the four-point function is the same whether it is computed using a conformal block expansion in the $s$- or in the $t$-channel as shown in figure \ref{fig.crossing4ptInt}.
\begin{figure}[tb]
	\centering
	\begin{tikzpicture}
	    \draw (4,1.6) -- (3,0.8) -- (3,-0.8) node[midway, right] {$k'$}-- (4,-1.6);
	    \draw (3,0.8) -- (2,1.6);
	    \draw (3,-0.8) -- (2,-1.6);
	    \node[right] at (4,1.6) {1}; 
	    \node[right] at (4,-1.6) {2}; 
	    \node[left] at (2,1.6) {1}; 
	    \node[left] at (2,-1.6) {2};
	    \node at (1.25,0) {\raisebox{-0.5cm}{$\sum\limits_{k'}$}};
        \node at (5,0) {$=0$};
	    \draw (-0.8,1) -- (-1.6,0) -- (-3.2,0) node[midway, above] {$k$}-- (-4,1);
	    \draw (-3.2,0) -- (-4,-1);
	    \draw (-1.6,0) -- (-0.8,-1);
	    \node[right] at (-0.8,1) {1}; 
	    \node[right] at (-0.8,-1) {2}; 
	    \node[left] at (-4,1) {1}; 
	    \node[left] at (-4,-1) {2};
	    \node at (0.25,0) {$-$};
	    \node at (-5,0) {\raisebox{-0.5cm}{$\sum\limits_{k}$}};
	\end{tikzpicture}
	\caption{Crossing equation for the 4-point function in a conformal field theory follows from the associativity of the OPE expansion and implies that the conformal block expansions given by expressions \eqref{eq.sblock} and \eqref{eq.tblock} should be convergent {\it and} equal.}
	\label{fig.crossing4ptInt}
\end{figure}

In an ensemble over approximate CFTs as given by \eqref{eq.avgobs}, requiring that the crossing equation is satisfied on ``average'' constrains the measure of the ensemble. However, as is the case for any statistical ensemble, also the variance is of primordial importance. The variance of the crossing equation is defined by
\begin{equation}
	\begin{tikzpicture}
	    \draw (-6,2) -- (5.5,2);
        \draw (-5.5,-1.75) .. controls (-6,-1) and (-6,1) .. (-5.5,1.75);
        \draw (4,1.6) -- (3,0.8) -- (3,-0.8) node[midway, right] {$k'$}-- (4,-1.6);
	    \draw (3,0.8) -- (2,1.6);
	    \draw (3,-0.8) -- (2,-1.6);
	    \node[right] at (4,1.6) {1}; 
	    \node[right] at (4,-1.6) {2}; 
	    \node[left] at (2,1.6) {1}; 
	    \node[left] at (2,-1.6) {2};
	    \node at (1.25,0) {\raisebox{-0.5cm}{$\sum\limits_{k'}$}};
	    \draw (-0.8,1) -- (-1.6,0) -- (-3.2,0) node[midway, above] {$k$}-- (-4,1);
	    \draw (-3.2,0) -- (-4,-1);
	    \draw (-1.6,0) -- (-0.8,-1);
	    \node[right] at (-0.8,1) {1}; 
	    \node[right] at (-0.8,-1) {2}; 
	    \node[left] at (-4,1) {1}; 
	    \node[left] at (-4,-1) {2};
	    \node at (0.25,0) {$-$};
	    \node at (-5,0) {\raisebox{-0.5cm}{$\sum\limits_{k}$}};
        \draw (5,-1.75) .. controls (5.5,-1) and (5.5,1) .. (5,1.75);
        \node[right] at (5.1,1.7) {$2$};
        \node[right] at (5.75,0) {$.$};
	\end{tikzpicture}
	\label{eq.Varcrossing4ptInt}
\end{equation}

The overline in the above expression represents the ensemble averaging over the approximate CFTs. This quantity is manifestly positive and measures the deviation from an ensemble of true CFTs, where crossing is satisfied in each member of the ensemble. Therefore, requiring that the variance of the crossing equation be small for a set of approximate CFTs imposes stricter restrictions on the CFT data. This can be incorporated in the measure that one uses to average over approximate CFTs. As we will demonstrate, this introduces non-Gaussian corrections for the statistical distribution over the CFT data in \eqref{eq.avgobs}.

We perform explicit computations in an ensemble relevant for pure gravity in AdS$_3$. This ensemble is specified through the moments of the dynamical data (in particular the OPE coefficients), rather than through the probability distribution over approximate CFTs. As mentioned above, this provides a second way to specify an ensemble of approximate CFTs.\footnote{up to subtleties that we discuss, for example the moment problem.} We start by computing the variance of the crossing equation in a purely Gaussian model as introduced in \cite{Chandra:2022bqq}, finding that the variance is large. We then introduce the appropriate non-Gaussianity (see \eqref{eq.4cnonGauss}) which significantly reduces the variance, forcing it to vanish at leading order in the large central charge limit, in accordance with our expectations for an ensemble of approximate CFTs.

\subsection{\texorpdfstring{A tensor model for AdS$_3$}{A tensor model for AdS3}}
In the third approach to approximate CFTs we explicitly construct a probability distribution that puts the majority of its weight on approximate solutions of the CFT constraints, in analogy with what was done in \cite{Jafferis:2022uhu,Jafferis:2022wez} for JT gravity with matter. The main idea is to first identify the random variables specifying the ensemble, in our case the set of operator dimensions $(h_i, \bar h_i)$ and the structure constants $C_{ijk}$. In the absence of further information the ensemble would be Gaussian, so as to maximise the entropy, but in fact we can feed more information into this model by identifying a set of constraints and exponentiating them into the potential. 

In the present context the constraints that are being exponentiated to define the potential $V$ are the modular covariance of torus one-point functions, as well as the four-point crossing of operators on the sphere. Note that this should ensure (approximate) modular and channel crossing invariance of all higher-point functions on surfaces of arbitrary genus, by invoking a generalization of the Moore-Seiberg construction \cite{Moore:1988qv} to the case of irrational CFTs. Very schematically, the model that we write down in Section \ref{sec.tensor} takes the form of an integral over a matrix $\Delta_{ij}$ and tensor $C_{ijk}$\footnote{The structure of our tensor model appears to be rather different than other tensor models that have been discussed in recent years (see \cite{Gurau:2019qag} for a review), but it would be interesting to investigate possible connections.}
\begin{equation}
    \int D[\Delta]D[C]\left(  \cdot \right)e^{-a V[\Delta, C]} \,,
\end{equation}
where insertions may be partition functions or correlation functions, expressed in terms of the basic `CFT data', $\left\{ \Delta_{ij}, C_{ijk} \right\}$. Naturally we think of the eigenvalues of $\Delta_{ij}$ as the conformal dimensions, while $C_{ijk}$ are the structure constants.
Note that such an exponentiation of constraints comes with a `large coefficient' $a$ in front of the squared constraint in the potential, which becomes a regulator in the model. We expect that the correct extrapolation to a continuum limit is an appropriate generalization of the the double-scaling procedure of \cite{Jafferis:2022uhu,Jafferis:2022wez}. As we explain in Section \ref{sec.tensor} this aspect requires further study, and it will be interesting to understand whether the continuum limit of our simplicial Virasoro gravity actually results in a continuous theory of AdS$_3$ gravity. The fact that the potential arises as the maximum entropy potential compatible with the operator content expected of the dual of pure gravity, namely the Viraosoro identity block and a continuum of states above the threshold at $(h, \bar h) \ge \frac{c-1}{24}$, strongly suggests such a connection should exist. An important aspect is that the random variables are the dimensions and OPE coefficients of all operators above the black-hole threshold, while the dimensions and OPE coefficients of light data (if present in the model) do enter the potential, but do not get integrated over. This is in line with the philosophy that we average over everything that is not fixed by low-energy supergravity, while holding the data accessible to low-energy supergravity fixed (up to the tolerance).

\subsubsection*{Plan of the paper}
While already implicit in the above arrangement of introductory sections, we nevertheless give a brief overview of the plan of the paper. We start with defining the notion of approximate CFTs in section \ref{sec.approxCFTs}. In section \ref{sec.var-cross} we study the higher moments (in particular the variance) of the crossing equation. We compute the non-Gaussian correction to the distribution of OPE coefficients, required to ensure that the variance of the crossing equation is obeyed to leading order in the large central charge limit.
Section \ref{sec.tensor} proposes a tensor model for a dual of pure 3d gravity in Anti-de Sitter space, to be thought of as a concrete probability measure on a space of approximate CFTs. We finally conclude with some remarks and open questions in section \ref{sec.discussion}.

\section{Approximate Conformal Field Theories \label{sec.approxCFTs}}

\subsection{Review of CFTs}

We will start by reviewing some general aspects of conformal field theories which will be useful for the rest of the paper, before introducing the concept of approximate CFTs. At the local level, a CFT$_d$ is fully specified by what is referred to as CFT data, namely the list of local operators with their scaling dimensions and spin as well as the OPE coefficients
\be \label{CFTdata}
\{\Delta_i, J_i ; C_{ijk}\} \,.
\ee
Here, $J_i$ denotes a representation of the rotation group $SO(d)$. A consistent conformal field theory is a complete list of such data which satisfies the fundamental axioms of conformal field theories: the theory should be unitary, causal, and respect crossing symmetry or associativity of the OPE. This highly constrains the allowed values for the CFT data \rref{CFTdata}, see \cite{Ferrara:1973yt,Polyakov:1974gs,Belavin:1984vu, Rattazzi:2008pe}. It is important that \textit{every} correlation function should obey crossing symmetry, \ie
\be
\braket{O_i(x_1) O_j(x_2) O_k(x_3) O_l(x_4)} \,,
\ee
should satisfy the crossing equations for all choices of operators $i,j,k$ and $l$. Higher-point correlation functions should also be crossing-invariant, although this follows from four-point crossing. In two-dimensions, CFTs must also satisfy modular covariance of one-point functions on the torus. Along with four-point crossing on the plane, it is believed that this provides sufficient information to define a consistent CFT, with well-defined crossing and modular-covariant $n$-point functions on arbitrary Riemann surfaces. This has been proven for rational CFTs \cite{Friedan:1986ua,Moore:1988uz,Moore:1988qv} but is believed to be true for all 2D CFTs. In higher dimensions, it is currently unknown what additional information is encoded on higher topologies, but it certainly involves non-local operators \cite{Belin:2018jtf}.

Crossing-symmetry has profound consequences on the CFT data. An efficient way to encode some of these constraints is through asymptotic formulas on the spectral or OPE densities. The best-known example of this is Cardy's formula which universally predicts the density of states of 2d CFTs \cite{Cardydos}. For OPE coefficients, this was first studied in \cite{Pappadopulo:2012jk} (see also \cite{Das:2017cnv,Cardy:2017qhl,Qiao:2017xif,Mukhametzhanov:2018zja,Collier:2019weq}). Consider for example the four-point function of identical operators
\be
\braket{O(x_1)O(x_2)O(x_3)O(x_4)} \,.
\ee
Now take an OPE limit between $x_1\to x_2$, $x_3\to x_4$. In this limit, in the $1\leftrightarrow 2$, $3\leftrightarrow 4$ channel, the identity operator dominates and produces a pole
\be
\braket{O(x_1)O(x_2)O(x_3)O(x_4)}  \sim \frac{1}{|x_{12}|^{2\Delta_O} |x_{34}|^{2\Delta_O}} +\cdots.
\ee

How is this pole reproduced in the cross-channel? It is easy to see that it cannot be reproduced by a single operator. Rather, the pole is reproduced by the collective contribution of many operators with $\Delta \gg 1$ with carefully tuned OPE coefficient. One finds the behaviour
\be \label{asymHHL}
\overline{C_{|OOO_H}|^2} \sim \frac{\Delta_H^{2\Delta_O-1}}{\rho(\Delta_H)} \,.
\ee
Here $\rho(\Delta_H)$ counts the total number of operators in a small window around $\Delta_H\gg1$, and $\overline{|C_{OOO_H}|^2}$ represents the average value of OPE coefficients squared for all operators in that window. For more details on the size of the window and corrections to this formula, see \cite{Qiao:2017xif,Mukhametzhanov:2018zja,Pal:2019zzr,Mukhametzhanov:2019pzy,Mukhametzhanov:2020swe,Das:2020uax,Pal:2023cgk}.

This shows that there must be fine-tuning between different sectors of the CFT data for crossing symmetry to be preserved. Naturally, \rref{asymHHL} only captures a tiny fraction of this fine-tuning (it is only sensitive to the identity operator in one channel), and only captures it on average in a sum over many heavy operators. This fine-tuning will be important to keep in mind in what follows.

\subsection{Defining approximate CFTs}

With this lightning review of CFTs in mind, we are ready to propose a definition for approximate CFTs. An approximate CFT is a list of data $\{\Delta_i, J_i ; C_{ijk}\} $ which approximately satisfies the constraints coming from the crossing equations. We will see that our framework is only consistent if we impose some conditions on the observables we allow ourselves to study. In particular, we will consider the following restrictions:
\begin{itemize}
\item  $n_{\max}$: we only consider correlation functions with at most $n_{\max}$ external operators.
\item $\Delta_{\max}$: we only allow correlation functions with external operators of scaling dimension at most $\Delta_{\max}$.\footnote{Other choices of restrictions are possible. For example, we can constrain the sum of dimensions of external operators, rather than constraining the total number and the maximal weight of the operators independently.} Note that this also caps off the maximal spin of external operators through unitarity. We emphasize that this cutoff doesn't restrict the scaling dimension of operators that are exchanged in the conformal block expansion of correlators. In other words, we don't restrict the size of the Hilbert space. We only restrict the scaling dimension of external operators, and internal operators can be arbitrarily heavy.
\item $\zl$: we constrain the allowed kinematics for the operators insertion $x_1, \cdots, x_{n_{\max}}$, in particular Lorentzian kinematics where operator insertions approach each other's lightcone. The Lorentzian limit which is most important to cut off is the Regge limit, and we introduce a limit on all possible cross-ratios $z_i$, namely that  $z_i>\zl \forall i$.
\item $g_{\max}$: for 2d CFTs, we only allow correlation functions on Riemann surfaces of genus at most $g_{max}$. We also restrict the modular parameters of the Riemann surfaces, similarly to how we restricted the kinematics for operators on the plane.
\end{itemize}

With these restrictions in place, we now have only a subset of all possible observables in a CFT. We call this set of observables $\{ \mathbb{O}_{\textrm{restr}}\}$. We impose crossing symmetry (and modular-covariance in $d=2$ or other relevant CFT constraints in general dimensions) for this reduced space of observables. In such a situation, we have fewer constraints on the CFT data than one would impose on an exact CFT. In particular, the data of operators $O_H$ with $\Delta_H>\Delta_{\max}$ is less constrained. Nevertheless, the heavy operators are still subject to the remaining CFT constraints on sub-threshold observables, as demonstrated by \eqref{asymHHL}, for example. This is because the heavy spectrum contributes to the sub-threshold observables as intermediate states. It is important to emphasize that the constraints always involve sums over the entire heavy spectrum (or a dense subset of it). Therefore, this only constrains the CFT data like $C_{LLH}, C_{HHH}$, averaged over suitable windows of $\Delta_H$ as described above. On the other hand, the OPE coefficients of individual operators are not strongly constrained. This means that individual violations of CFT constraints for the heavy operators can be big, but only if these violations are distributed across the entire heavy spectrum in a correlated fashion, such that the total violation is very small on average. We illustrate this point in an explicit example in Section \ref{sec.example-approxCFT} below.

Quite interestingly, we can deduce from this chain of arguments that requiring the exact CFT constraints for the restricted set of observables, $\{\mathbb O_{\rm restr}\}$,
\be
\textrm{Crossing} = 0 =  \textrm{Modular Covariance }  \,, \forall \ \mathbb{O} \in  \{ \mathbb{O}_{\textrm{restr}}\}
\ee
is too restrictive and most likely inconsistent. For example, even if the four-point function of some light operator $\mathbb O$ were exactly crossing-invariant, its eight-point function would involve (as an intermediate step through recursive application of the operator product expansion) a sum of four-point functions with operators $\mathbb O_H$. But we have already argued above that the four-point function of $\mathbb O_H$ obeys the CFT constraints only in an averaged sense, and not exactly. Thus, the sub-threshold observables also satisfy the CFT constraints only up to the `statistical error' that is inherited from the high-energy observables. Therefore, we need to introduce a tolerance parameter $\mathbb{T}$ which is governed by the error incurred by the CFT constraints in the high energy spectrum. We take this tolerance to be small. Therefore we impose\footnote{It may be possible to consider some subset of restricted observables, and demand them to be exactly crossing invariant. Because we fear it is not consistent to demand that all restricted observables exactly satisfy crossing, we will take the more conservative route to introduce a tolerance for the entire $\mathbb{O}_{\textrm{restr}}$, even if this can perhaps be improved upon.}
\be
\textrm{Crossing}, \textrm{Modular Covariance } < \mathbb{T}  \,, \forall \ \mathbb{O} \in  \{ \mathbb{O}_{\textrm{restr}}\}~.
\ee
It is important to remember that the CFT constraints like crossing or modular equations are functions of kinematic/modular parameters, up to the restrictions imposed by $z_{\min}$, etc. We demand that the constraints are satisfied up to the tolerance uniformly through the allowed space of kinematics. 
We define an approximate CFT $\tilde{\mathcal{C}}$ as any list of data that corresponds to a choice of 
\be
\tilde{\mathcal{C}} \longleftrightarrow \{ \mathbb{O}_{\textrm{restr}},\mathbb{T}\} \,.
\ee

Note that in many instances (excluding the cases of conformal manifolds), imposing the full set of constraints uniquely fixes a CFT: there is a unique solution to the bootstrap equations. Here, the reduced set of constraints may not allow us to find a unique solution, but rather fixes the data up to some allowed subspace of the full parameter space. This is in fact a desired aspect of our construction that we will use momentarily. We will sometimes refer to this subspace as an \emph{island}, in reference to what is obtained in the numerical bootstrap of the 3D Ising model, but the topology of the subspace may be extremely complicated. 

Our notion of approximate CFTs can also be viewed from a Wilsonian perspective, where only low-energy observables can be accessed by a physical observer and thereby should be the only ones to be reliably constrained. Here, low-energy is meant in the sense of low scaling dimension, which does map to low-energy on the sphere. The fact that this is sufficient to constrain some parts of the high energy spectrum is reminiscent of dispersion relations in S-matrix theory, see \cite{Hartman:2022zik} for an overview. Our approach also naturally connects to the concept of coarse-graining: the correlation functions of the operators in the high energy spectrum must also obey the CFT constraints, but only in an \emph{averaged} or \emph{coarse-grained} sense. This notion of coarse-graining is similar to one that is used to understand quantum ergodicity and ETH in quantum theories, \cite{DAlessio:2015qtq, Altland:2020ccq}. This connection can be implemented efficiently in a 2d CFT: while we do not impose the crossing invariance of individual heavy operators within our formulation of approximate CFTs, requiring the approximate modular invariance of the partition function on genus-two surfaces implies that the crossing equation is satisfied approximately for the heavy operators when a sum over many heavy external operators is taken.

Finally, we should mention that the observables of an approximate CFT are partition functions or correlation functions of at most $n_{\max}$ operators of scaling dimension no greater than $\Delta_{\max}$. In general, these observables are not single-valued: they explicitly depend on a channel decomposition. Therefore, the observables are really correlation functions with a specification of channel decomposition. Alternatively, one can say that correlation functions are true observables if we only want to specify them up to some precision fixed by the tolerance parameter.

\subsection{Inspiration from Holography}

The definition of approximate CFTs we have given above can be applied in principle to any context, going from the 3d Ising model to $\mathcal{N}=4$ SYM. However, it finds a particularly interesting application within the holographic setup. Bulk calculations are performed using the low-energy effective action, namely that of Einstein gravity or supergravity. Holography allows us to compute certain observables of the dual CFT to good approximation, but not all of them. For example, we can consider a four-point function of supergravity modes on the plane, or we can consider a thermal partition function which will be dual to a black hole geometry. 
However, we cannot reliably use the supergravity action to compute an $N^2$-point function of the stress-tensor in $\mathcal{N}=4$ SYM, for example. Even though gravity is weakly coupled, it becomes effectively strongly-coupled when we probe it with this many operators. Similarly, while we can compute correlation functions in deep Lorenztian kinematics (i.e. in the Regge limit for timescales beyond the scrambling time) using Einstein gravity, we should not trust them if the scattering energy becomes too big. In particular, if the energy becomes too big, the scattering process will create black holes, which can take us outside of the regime of validity of effective field theory.
Other observables are even more problematic. Consider four operators that create black hole microstates, and let us study their four-point function. This observable cannot be studied within supergravity. Contrary to the $N^2$-point function, it is now a problem of principle, meaning the corresponding correlation function cannot even be defined using just low-energy supergravity. It is not merely a technical problem of computing the correlator in practice. 

This provides us  with a natural proposal for the value (or rather $N$-scaling) of the parameters that define an approximate CFT in the context of holography. We define the parameter $N$ to be the stress-tensor two-point function $\braket{TT}=N$, and the relation to Newton's constant is given by\footnote{Note that this is not the standard meaning of $N$ in $\mathcal{N}=4$ SYM, for which $N\to N^2$.}
\be
N\sim \frac{\ell^{d-1}}{G_N} \,.
\ee
We expect the following scaling of the parameters
\be
n_{\max} \ll N \,, \qquad \Delta_{\max} \ll N \,, \qquad \zl \ll \log N \,, \qquad g_{\max} \ll N \,.
\ee
The first two scalings have been argued above and are related to the breakdown of the weak coupling of gravity, as well as correlation functions of external black hole microstates. The condition on $\zl$ can also be seen as from the breakdown of the perturbative regime of gravity in the Regge limit. The value $\log N$ for the cross-ratio maps to Lorentzian times of order the scrambling time, or of center of mass energies in the AdS scattering process of order the Planck mass (see \cite{Maldacena:2015waa,Roberts:2014ifa}). At those timescales, we exit the perturbative regime of gravity.

Finally, for the genus of the Riemann surface in 2d CFTs, we do not have a direct argument. It would be interesting to understand better the breakdown of EFT on complicated Riemann surfaces, but without a more detailed knowledge, it seems plausible to consider $g_{\max} \ll N$. One way to motivate this is based on the observation that a genus $g$ manifold can be constructed from a $2n$-point function on the sphere by plumbing pairs of states. Thus it is plausible that the scaling of maximum allowed genera should be the same as that of the number of insertions in a correlation functions.

It is worth noting that computations done in Euclidean gravity always seem to satisfy crossing or modular covariance. This typically follows from summing over topologies or summing over channels in Witten diagrams. Naturally, since Einstein gravity is only an EFT, we do not trust our calculations to arbitrary precision, in particular we can at best claim to compute the observables up to corrections of the order
\be
e^{-\frac{1}{G_N}} \sim e^{-N} \,.
\ee
This suggests that the tolerance parameter for holographic CFTs should be of the order of our ignorance in observables, which should be taken to be exponentially small in the central charge
\be \label{scalingtolerance}
\mathbb{T}\sim e^{-N} \,.
\ee

\subsection{Example of an Approximate CFT}\label{sec.example-approxCFT}
With the definitions of the previous section in hand, it is important to check whether the type of object we are defining actually exists, or whether we are asking for too much and the only type of data that satisfy the rules of an approximate CFT are those of an actual CFT. We will see that it is actually straightforward to produce approximate CFTs that are not true CFTs. In what follows, we will generate approximate CFTs by deforming the CFT data of a consistent CFT by a small amount and requiring a relaxed set of CFT constraints as described above. Therefore, within the island of approximate CFTs that emerges, there is always a true CFT. It is possible that islands of approximate CFTs exist such that there are no true CFTs within them. This is an interesting question for holography, but we do not address it in the  present work.\footnote{It is particularly interesting in light of the swampland program. If there existed islands of approximate holographic CFTs that do not contain a true CFT, it would imply that the bulk gravitational EFT is not UV-completable and is at best dual to an ensemble of approximate holographic CFTs.}

The most natural guess of what an approximate CFT could be, is to start with a true CFT and simply change the dimension of one given operator by a small amount. We will take an operator $O_0$, with $\Delta_{0}\gg1$, and shift
\be
\Delta_0\to\Delta_0+\epsilon \,.
\ee
At very first glance, we might think the new data satisfy our definition of an approximate CFT. This turns out not to be true, as can be seen from the discussion around \rref{asymHHL}: if we change only the weight of the operator $O_0$ and nothing else, then the correlation functions (in particular the four-point function) of this operator fail to satisfy the crossing equation. If we take an OPE limit of the correlation function $\braket{O_0O_0O_0O_0}$, in the channel where the identity dominates we will find
\be
\braket{O_0(x_1)O_0(x_2)O_0(x_3)O_0(x_4)}  \sim \frac{1}{|x_{12}|^{2\Delta_O+2\epsilon} |x_{34}|^{2\Delta_O+2\epsilon}} \,.
\ee
In the cross-channel, recall that the pole is reproduced by the sum over the heavy tail of operators at very large scaling dimension (much greater than $\Delta_0$ itself). But these have not been altered, so we find
\be
\braket{O_0(x_1)O_0(x_2)O_0(x_3)O_0(x_4)}  \sim \frac{1}{|x_{12}|^{2\Delta_O} |x_{34}|^{2\Delta_O}} \,.
\ee

One easily sees that the ratio of the two expressions, which must equal to one in a crossing invariant theory, is unbounded in the OPE limit. So we obtain an unbounded violation of the crossing equation. This problem is easily circumvented, simply by requiring that
\be
\Delta_{\max} < \Delta_0 \,,
\ee
in which case the correlation function $\braket{O_0(x_1)O_0(x_2)O_0(x_3)O_0(x_4)}$ does not lie within the set of observables we are allowed to consider.\footnote{We could also have fixed crossing by simultaneously changing the OPE coefficients $C_{O_0O_0O_H}$ in a coordinated way, but we here we try to construct approximate CFTs with the minimal number of modifications of the starting CFT.}

However, many other observables also receive contributions from the high energy part of the spectrum. For example, this is true of correlation functions of light external operators. Let us consider the example of a 4-point function of a light operator $O_L$ whose weight $\Delta_L \ll \Delta_{max}$, and which receives a contribution from the exchange of the heavy operator $O_0$ (recall its dimension is perturbed as $\Delta_0 \to \Delta_0 + \epsilon$). We will consider the case of $d=2$, but similar considerations hold in higher dimensions as well. The conformal block expansion of the 4-point function can be written as,
\be
\braket{O_L(z_1)O_L(z_2)O_L(z_3)O_L(z_4)}= \frac{1}{|z_{12}|^{2\Delta_L} |z_{34}|^{2\Delta_L}} \sum_O | C_{O_LO_LO}|^2 |z|^{2h_O} |F(h_O,h_O,2h_O,z) |^2  \,,
\ee
where $F$ is the hypergeometric function, $h_O=\bar{h_O}=\Delta/2$ for a scalar operator, and we have used the cross-ratio
\be
z=\frac{z_{12}z_{34}}{z_{14}z_{23}} \,.
\ee
The difference between the original CFT correlator and the correlator computed with the modified data is given by a difference of two conformal blocks
\be
\frac{|C_{O_LO_LO_0}|^2}{|z_{12}|^{2\Delta_L} |z_{34}|^{2\Delta_L}} \left(|z|^{2h_O}|F(h_0,h_0,2h_0,z)|^2 - |z|^{2(h_O+\epsilon)}|F(h_0+2\epsilon,h_0+2\epsilon,2h_0+4\epsilon,z)|^2 \right) \,.
\ee
The quantity above transforms non-trivially under a crossing transformation, and the new
data thus fails to satisfy the crossing equation.
Moreover, the violation of the crossing constraint due to the above expression is \emph{not} uniformly bounded in $z$. The hypergeometric function has three singular points, 0, 1 and $\infty$. It is bounded at zero and infinity, but diverges logarithmically as $z\to 1$. In the limit $z\to1$, the conformal block behaves as
\be
F(h_O,h_O,2h_O,z)\sim -\frac{\Gamma(2h_O)}{\Gamma(h_O)^2} \log (1-z) \,.
\ee
The size of the violation of crossing invariance can be estimated in this limit, where for $h_0\gg1, z\to1$, together with $\epsilon \ll 1$ we have\footnote{The proportionality constant is an $\mathcal O(1)$ number and hence not important for our purpose.}
\begin{equation}\label{violationcrossing1op}
    \text{violation of crossing } \sim \epsilon \times 4^{h_0} \sqrt{2 h_0} \, \log(1-z)~.
\end{equation}
Note that the violation of the crossing equation is indeed controlled by $\epsilon$ but nevertheless, for sufficiently small values of $1-z$, it is still large. Since this violation is unbounded in the $z\to1$ limit, we have not succeeded in finding an approximate CFT.  
However, this problem can easily be fixed by changing the dimensions of multiple operators in a correlated fashion. Let us show this by deforming the dimensions of two different operators $\mathcal O_1$ and $\mathcal O_2$,
\be
h_1\to h_1+\epsilon_1 \,, \qquad h_2\to h_2+\epsilon_2 ~.
\ee 
The violation of the crossing equation is now proportional to
\be
\left(|C_{LLO_1}|^2\left(-\frac{\Gamma(2h_1)}{\Gamma(h_1)^2} + \frac{\Gamma(2h_1+2\epsilon_1)}{\Gamma(h_1+\epsilon_1)^2}\right)-|C_{LLO_2}|^2\left(-\frac{\Gamma(2h_2)}{\Gamma(h_2)^2} + \frac{\Gamma(2h_2+2\epsilon_2)}{\Gamma(h_2+\epsilon_2)^2}\right) \right)\log (1-z) \,.
\ee
We can now simply correlate our choice of $\epsilon_{1,2}$ to set the coefficient of $\log(1-z)$ to zero. One can see that this now produces an approximate CFT. The crossing equation is violated, but the violation is bounded in (Euclidean) cross-ratio space.\footnote{It is most likely badly violated for Lorentzian kinematics, but this problem is controlled by the limitation of Lorentzian kinematics we introduced.} One can quickly generalize this procedure by displacing $n$ operators that lie above $\Delta_{\max}$ as long as
\be
\sum_{i=1}^n |C_{LLO_i}|^2\left( -\frac{\Gamma(2h_i)}{\Gamma(h_i)^2} + \frac{\Gamma(2h_i+2\epsilon_i)}{\Gamma(h_i+\epsilon_i)^2} \right)=0 \,.
\ee
Thus far, we have held the OPE coefficients fixed but it is straightforward to obtain even more examples of approximate CFTs by changing OPE coefficients as well, in which case the condition becomes
\begin{equation} \label{conditionmanyops}
\sum_{i=1}^n \left(-|C_{LLO_i}|^2\frac{\Gamma(2h_i)}{\Gamma(h_i)^2} + |C'_{LLO_i}|^2\frac{\Gamma(2h_i+2\epsilon_i)}{\Gamma(h_i+\epsilon_i)^2} \right) =0 \,.
\end{equation}

Not only have we just seen that approximate CFTs actually exist, but there are is a huge number of them. This provides a landscape of theories to average over, as we now describe.

\subsection{Averaging over Approximate CFTs}\label{sec.AveragingApproxCFT}

As we have already emphasized, one of the principal reasons to introduce the notion of approximate CFT is that this structure gives us the individual elements of an appropriate  ``ensemble of CFTs''. In general, CFTs are isolated points in parameter space, and even in the rare case of a conformal manifold, the number of parameters (i.e. the dimension of the conformal manifold) tends to be small, and certainly much smaller than the number of degrees of freedom in the case of holographic CFTs. On the other hand, approximate CFTs offer a huge landscape of individual elements which all obey the same conditions, which can be seen for example in equation \rref{conditionmanyops}. We will call the space of approximate CFTs $\mathcal{A}$.

This ensemble of individual theories can now be averaged over. As we have already discussed, the motivation for such averages comes from the need to efficiently describe chaotic CFTs, and understand the relevant ensemble from which an individual CFT is drawn. Another motivation is to identify the right coarse-graining procedure which is relevant for low-energy approximations in gravity, which notably give rise to connected multi-boundary solutions of the wormhole type. We will come back to gravity shortly.

An averaged CFT is then a set of CFT data $\left\{ \Delta_i, J_i, C^k_{ij}\right\}$, together with a joint probability density $P\left( \left\{ \Delta_i, J_i, C^k_{ij}\right\} \right)$ for the scaling dimensions (at each spin) and OPE coefficients.
 From this probability density, we can consider the averaged OPE
\begin{equation}   
\overline{{\cal O}_i(x) {\cal O}_j(0)}  \equiv \int_{\mathcal{A}} P\left( \left\{ \Delta_k, J_k, C^k_{ij}\right\}\right) \int d\Delta_k  \rho(\Delta_k) C^k_{ij}(x) {\cal O}_k \equiv \int d\Delta_k \overline{ \rho(\Delta_k) C^k_{ij} (x)} {\cal O}_k \,.
\end{equation}
We can also study the averaged 4-point crossing equation, which reads
\begin{equation}\label{eq.averageCrossing}
\overline{\cal CR}(u,v) := \overline{\sum_{\Delta, {\ell}} p_{\Delta, \ell} F_{\Delta, \ell} (u,v) } \,,
\end{equation}
where as usual the $p_{\Delta, \ell}$ are positive coefficients built from (squared) OPE coefficients and
\begin{equation}
F_{\Delta, \ell} (u,v) := u^\Delta g_{\Delta, \ell}(u,v) - v^\Delta g_{\Delta, \ell} (v,u) \,,
\end{equation}
is expressed in terms of the global conformal blocks $g_{\Delta, \ell}(u,v)$. 

Since the individual approximate CFTs we have averaged over had a bounded violation of crossing and the probability density integrates to unity, the averaged crossing equation is approximately zero, and bounded by the tolerance parameter
\be
|\overline{{\cal CR}(u,v)}| < \mathbb{T} \,.
\ee
In some cases, it may be possible to find a crossing equation that vanishes exactly upon averaging, although we do not require this. 

Satisfaction of the bootstrap constraints on average, whether approximately or exactly, is not enough to characterise the ensemble. Instead, the higher statistical moments of the bootstrap equations need to be taken into account as well. For example, the variance of the 4-point crossing equation
\begin{equation}
\sigma_{(2)} \left[ {\cal CR} (u,v)\right] := \overline{\sum_{\Delta,\ell} \sum_{\Delta', \ell'} p_{\Delta, \ell} p_{\Delta', \ell'} F_{\Delta,\ell}(u,v) F_{\Delta', \ell'}(u,v) } - \overline{\cal CR}^2\,,
\end{equation}
is of crucial importance. It identically vanishes only for averages over exact CFTs, and since it is an integral of positive quantities over a positive probability density, it is positive definite. It is the simplest diagnostic of the deviation from having an ensemble of exact CFTs. Similarly, one can define the higher moments $\sigma_{(n)} \left[ {\cal CR}(u,v) \right]$. It should be clear that these higher moments of the bootstrap equations tightly constrain the connected higher-point moments of the distribution of $\left\{ \Delta_i, J_i,C^k_{ij} \right\}$, and as we will see, require non-Gaussianities in the statistics of OPE coefficients. The tolerance parameter controls the hierarchy between the moments of the crossing equation, as 
\begin{eqnarray}\label{eq.HierarchyBootstrapConstraints}
\overline{{\cal CR}(u,v)} &<& \mathbb{T} \nonumber\\
\sigma_{(2)} \left[ {\cal CR}(u,v) \right] &<& \mathbb{T}^2 \nonumber\\
&\vdots& \nonumber\\
 \sigma_{(n)} \left[ {\cal CR}(u,v) \right] &<& \mathbb{T}^n \,.
\end{eqnarray}
For holographic CFTs, the scaling \rref{scalingtolerance} suggests higher moments are further and further exponentially suppressed in $N$.

Finally, note that there is some flexibility in the choice of probability distribution, in particular in terms of the data below $\Delta_{\max}$. We expect that in general, the data below $\Delta_{\max}$ is not fully fixed in a family of approximate CFTs.\footnote{In holography for example, we expect the supergravity data to get modified by black hole physics, but only non-pertubatively in $N$. This would affect the renormalization of the mass of non-BPS supergravity fields, or affect the binding energy of certain multi-particle states.} Therefore, one could take a probability measure that allows the light data to change over the ensemble. In practice however, it may be more convenient to fix this data exactly. We will mostly be thinking about probability distributions of this type.

\subsubsection*{Specifying the ensemble vs specifying the moments}

It is worthwhile to discuss the distinction between specifying an ensemble and specifying the moments. One may specify an ensemble through its moments, and from a practical point of view, specifying the low-lying moments is sufficient to compute simple expectation values and vice-versa. There are however two  conceptual issues that could arise with such an approach. First, it is not clear whether a set of chosen moments leads to a consistent ensemble. There are infinitely many conditions on the moments that need to be satisfied for the ensemble to be consistent. For example, the simplest of these conditions is that the variance is positive, but they quickly become complicated. 
Second, the ensemble can only be specified if the tail of the distribution, namely the asymptotic behaviour of the high moments is well controlled.\footnote{A generating function for the distribution of moments was defined in \cite{Belin:2021ryy}, and in principle a dictionary can be constructed between the probability distribution and the generating function. This will work well for the first few moments, but the tail of the distribution and its connection to the very high moments is subtle to understand.} In general, this is very delicate to probe. For the moments of the OPE coefficients, this involves controlling partition functions on Riemann surfaces of arbitrarily complicated topology and it is far from clear that this can be achieved. In some channels, this was accomplished in \cite{Belin:2021ryy} but the number of channels grows with the genus of the Riemann surface, which makes tracking the tail of the distribution extremely difficult.

\subsubsection*{Connection with semi-classical gravity}

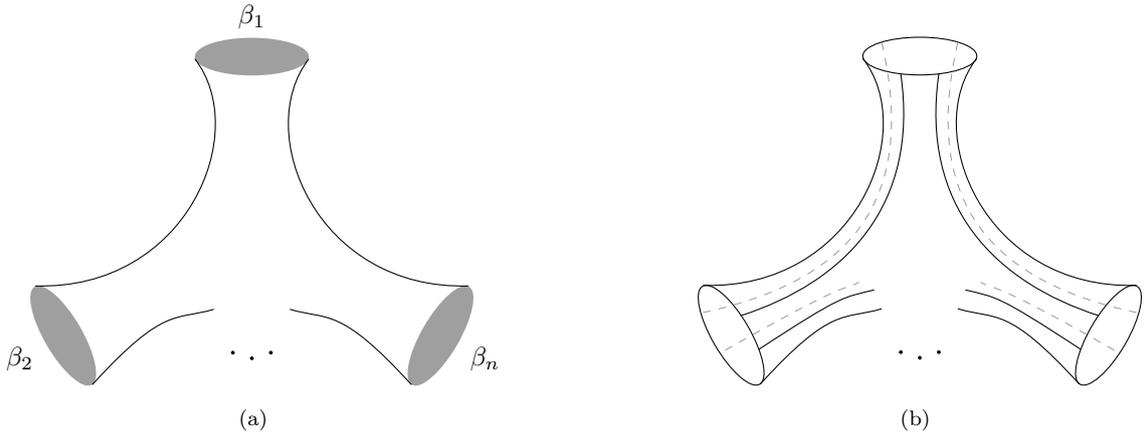
\begin{figure}[tb]
    \centering
	\subfigure[]
    {\begin{tikzpicture}
	    \fill[color=gray!75] (0,4) ellipse (0.75 and 0.25);
        \node[above] at (0,4.25) {$\beta_1$};
        \fill[rotate around={-60:(0,0)}, color=gray!75] (-1.5,-2) ellipse (0.75 and 0.25);
        \node[left] at (-2.75,0) {$\beta_2$};
        \fill[rotate around={60:(0,0)}, color=gray!75] (1.5,-2) ellipse (0.75 and 0.25);
        \node[right] at (2.75,0) {$\beta_n$};
        \draw (0.75,3.97) ..controls (0,3) and (0.85,0.9) .. (2.85,0.96);
        \draw (-0.75,3.97) ..controls (0,3) and (-0.85,0.9) .. (-2.85,0.96);
        \draw (2.1,-0.35) .. controls (1.15,0.7) and (1,0.5) .. (0.5,0.65);
        \draw (-2.1,-0.35) .. controls (-1.15,0.7) and (-1,0.5) .. (-0.5,0.65);
        \fill[color=black] (0,0) circle(0.025);
        \fill[color=black] (0.25,0.075) circle(0.025);
        \fill[color=black] (-0.25,0.075) circle(0.025);
	\end{tikzpicture}
    }\hfill
    \subfigure[]
    {\begin{tikzpicture}
	    \draw (0,4) ellipse (0.75 and 0.25);
        \draw[rotate around={-60:(0,0)}] (-1.5,-2) ellipse (0.75 and 0.25);
        \draw[rotate around={60:(0,0)}] (1.5,-2) ellipse (0.75 and 0.25);
        \draw (0.75,3.97) ..controls (0,3) and (0.85,0.9) .. (2.85,0.96);
        \draw (-0.75,3.97) ..controls (0,3) and (-0.85,0.9) .. (-2.85,0.96);
        \draw (2.1,-0.35) .. controls (1.15,0.7) and (1,0.5) .. (0.5,0.65);
        \draw (-2.1,-0.35) .. controls (-1.15,0.7) and (-1,0.5) .. (-0.5,0.65);
        \fill[color=black] (0,0) circle(0.025);
        \fill[color=black] (0.25,0.075) circle(0.025);
        \fill[color=black] (-0.25,0.075) circle(0.025);
        \draw[dashed, color=gray!75] (-.5,4.18) ..controls (0,2) and (-1,1) .. (-2.85,0.6);
        \draw[dashed, color=gray!75] (0.5,4.18) ..controls (0,2) and (1,1) .. (2.85,0.6);
        \draw (-.25,3.76) ..controls (0,2) and (-1,1) .. (-2.35,0.56);
        \draw (.25,3.76) ..controls (0,2) and (1,1) .. (2.35,0.56);
        \draw (-2.12,0.11) ..controls (-1,0.85) and (-1,0.8) .. (-.6,0.9);
        \draw (2.12,0.11) ..controls (1,0.85) and (1,0.8) .. (.6,0.9);
        \draw[dashed, color=gray!75] (-2.56,0.1) ..controls (-1,1) and (-1,0.9) .. (-.8,1.);
        \draw[dashed, color=gray!75] (2.56,0.1) ..controls (1,1) and (1,0.9) .. (.8,1.);
	\end{tikzpicture}}
    
    \caption{(a) The connected spectral correlation functions that arise in the field theory are captured by  multi-boundary wormholes in the gravitational dual. We show here `fixed-length' boundary conditions, fixing the inverse temperatures at each disconnected component of the boundary. The energy correlations of \eqref{eq.energyCorrs} are obtained by inverse Laplace transform. (b) The same geometries also contribute to connected multi-boundary correlation functions, where the boundary conditions include operator insertions.}
    \label{fig:my_label}
\end{figure}

One of the goals of the averaging over approximate CFTs is to make contact with semi-classical gravity. Wormhole geometries, as depicted in Fig. \ref{fig:my_label}, control the connected higher-point moments of either the spectral densities
\begin{equation} \label{eq.energyCorrs}
\overline{ \rho(E_1) \cdots \rho (E_n)}\Bigr|_{\rm c}\,,
\end{equation}
or the OPE coefficients
\begin{equation}
\overline{ C_{i_1 j_1 k_1} \cdots C_{i_n j_n k_n}}\Bigr|_{\rm c}\,.
\end{equation}
These moments can be extracted from multi-boundary correlation functions (on arbitrary topology and with or without operator insertions) of the type
\begin{equation}
\overline{\left\langle {\cal O}(x_1) \cdots {\cal O}(x_n) \right \rangle  \cdots \left\langle {\cal O}(x_1) \cdots {\cal O}(x_n) \right \rangle}\Bigr|_{\rm c}\,.
\end{equation}

The gravitational theory does not tell us whether the wormholes come from coarse-graining or from an explicit average, but as we have already discussed several times, the distinction is not so important here. In either case, we propose that the correct probability distribution (corresponding to an effective ensemble averaging or an explicit one) should be extracted by matching with bulk computations involving wormholes. Techniques for defining such probability distributions, in the context of two-dimensional JT gravity were recently described in \cite{Jafferis:2022uhu,Jafferis:2022wez}, and we will describe a version applicable to 2d CFT in section \ref{sec.tensor}. 

Note that in our ensemble, the 4-point crossing constraint \eqref{eq.averageCrossing} is only approximate. It is natural to ask whether 4-point crossing (or modular covariance) is only approximately satisfied in gravitational computations or not. It is hard to give an exact answer to this question without making some assumptions, and we discuss this further in the discussion. Nevertheless, a gravitational effective field theory cannot compute \textit{any observable} to an accuracy better than $e^{-N}$, and hence it is not meaningful to ask whether exact crossing is imposed by semi-classical gravity or not.

\section{The moments of crossing and essential non-Gaussianity}\label{sec.var-cross} 
We are now interested in computing quantities in the ensemble of approximate CFTs, in particular correlation functions of operators below $\Delta_{\max}$ or partition functions. For a given observable, it is interesting to quantify its deviation from being a true CFT observable in a candidate ensemble. As a first test, one could try to study the average of the crossing equation (various types of crossing equations could be studied, either four-point crossing or modular crossing on higher genus surfaces in $d=2$). This is not a very stringent test though, as the crossing equation is not sign-definite, so it may average to zero over the ensemble while still having large fluctuations. It is thus more natural to study the next moment, namely the square (or variance) of the crossing equation. For the crossing of a four-point function, this would read
\begin{figure}[h!]
	\centering
	\begin{tikzpicture}
	    \draw (-6,2) -- (5.5,2);
        \draw (-5.5,-1.75) .. controls (-6,-1) and (-6,1) .. (-5.5,1.75);
        \draw (4,1.6) -- (3,0.8) -- (3,-0.8) node[midway, right] {$k'$}-- (4,-1.6);
	    \draw (3,0.8) -- (2,1.6);
	    \draw (3,-0.8) -- (2,-1.6);
	    \node[right] at (4,1.6) {1}; 
	    \node[right] at (4,-1.6) {2}; 
	    \node[left] at (2,1.6) {1}; 
	    \node[left] at (2,-1.6) {2};
	    \node at (1.25,0) {\raisebox{-0.5cm}{$\sum\limits_{k'}$}};
	    \draw (-0.8,1) -- (-1.6,0) -- (-3.2,0) node[midway, above] {$k$}-- (-4,1);
	    \draw (-3.2,0) -- (-4,-1);
	    \draw (-1.6,0) -- (-0.8,-1);
	    \node[right] at (-0.8,1) {1}; 
	    \node[right] at (-0.8,-1) {2}; 
	    \node[left] at (-4,1) {1}; 
	    \node[left] at (-4,-1) {2};
	    \node at (0.25,0) {$-$};
	    \node at (-5,0) {\raisebox{-0.5cm}{$\sum\limits_{k}$}};
        \draw (5,-1.75) .. controls (5.5,-1) and (5.5,1) .. (5,1.75);
        \node[right] at (5.1,1.7) {$2$};
        \node[right] at (5.75,0) {$.$};
	\end{tikzpicture}
	\label{fig.Varcrossing4ptInt2}
\end{figure}
\newline
The variance is the first strictly positive moment, so it will be of particular interest to us. Being strictly positive, it cannot average out over the ensemble and provides a good estimate to the deviation from the ensemble being a true CFT. Higher moments are interesting as well, and we will comment on them at the end of this section.

 In the rest of the section, we will study this variance in an ensemble relevant for three-dimensional gravity. We will start by reviewing the ensemble as presented in \cite{Chandra:2022bqq}, which is a Gaussian ensemble for OPE coefficients. While this ensemble is built such that the mean of various crossing equations vanishes in the semi-classical limit, the variance of these equations do not vanish and in fact give rise to large deviations. We will then proceed to add the appropriate non-Gaussianities in order to ``fix" the variance of the crossing equation.\footnote{Non-Gaussian corrections to the ensemble were already discussed in \cite{Chandra:2022bqq}, but only in cases where they lead to small corrections in the semi-classical limit. Here the non-Gaussianities will be leading order effects.} It is important to note that non-Gaussianities for the distribution of OPE coefficients in any individual CFT always exist, and are in fact necessary to satisfy the crossing equations for higher-point correlation functions (or modular invariance on higher genus surfaces) \cite{Belin:2021ryy,Anous:2021caj}. What we will see here, is that the same non-Gaussianities are also necessary to enforce small higher moments of the crossing equation for lower-point functions.

\subsection{A Gaussian Ensemble for Pure Gravity in \texorpdfstring{AdS$_3$}{AdS3}?}

We will study the variance of the crossing equation in an ensemble of approximate CFTs relevant for pure three-dimensional gravity with or without conical defects, introduced in \cite{Chandra:2022bqq}. It will turn out that studying the square of crossing forces us to introduce non-Gaussiantities. To start with, let us specify the ensemble of  \cite{Chandra:2022bqq}. It contains no light primary operators (i.e. operators with dimension $\Delta\sim c^0$) other than the identity operator. The quadratic ensemble is specified through the moments of the dynamical data in the following way:
\begin{itemize}
\item Other than the identity operator, the high-energy part of the density of states is given by Cardy's formula \cite{cardyformula}
\be \label{cardy}
\overline{\rho(h,\bar{h})}\approx \rho_0(h) \rho_0(\bar{h})= e^{2\pi \sqrt{\frac{c}{6}\left(h-\frac{c}{24}\right)}}e^{2\pi \sqrt{\frac{c}{6}\left(\bar{h}-\frac{c}{24}\right)}} \,,
\ee
which extends down to from infinity to $h=c/24$. Note that the density of states is a continuous function of spin, $s = h-\bar{h}$, such that integrality of spin is not enforced in this ensemble. Moreover, the density of states is considered to be essentially ``classical" within the ensemble, such that higher point moments of $\rho(h,\bar{h})$ factorize exactly into one-point functions given by \rref{cardy}.

\item In addition to these operators which form a continuous spectrum, one can include a discrete set of operators with $h/c$ fixed and smaller than $1/24$ in the large $c$ limit. These operators are dual to very massive particles that backreact to produce conical defects in the dual bulk theory.

\item It remains to describe the statistics of the OPE coefficients. The distribution of OPE coefficients is given by a Gaussian distribution, such that
\be \label{Gaussianmoment}
\overline{C_{ijk}C_{lmn}^*} = C_0(h_i,h_j,h_k)C_0(\bar{h}_i,\bar{h}_j,\bar{h}_k)(\delta_{i,l}\delta_{j,m}\delta_{k,n}+\text{signed perm}) \,,
\ee
where ``signed perm'' means permutations of the indices with a sign fixed by the sum of spins and the function $C_0$ is a known function (given below in \eqref{eq.C0formula} in Liouville notation) which is related to the DOZZ formula of Liouville theory, and is obtained by studying the constraints of modular invariance at genus-two in a 2d CFT, see \cite{Cardy:2017qhl,Collier:2019weq,Chandra:2022bqq} for more details.\footnote{We will often use the compact notation $|C_0|^2$ for either the $C_0$ function, the density of states, or other functions of holomorphic weights. These should be understood as $|C_0|^2=C_0(h)C_0(\bar{h})$, and not as $C_0(h)C_0(h)^*.$}

For the sake of completeness, and because we will make much use of it in later sections, we quote here the $C_0$-formula referred to in equation \eqref{Gaussianmoment}, as derived in \cite{Ponsot:1999uf, Ponsot:2000mt} and used recently in \cite{Collier:2019weq,Chandra:2022bqq}
\begin{equation}\label{eq.C0formula}
    C_0 (P_i, P_j, P_k) = \frac{1}{\sqrt{2}} \frac{\Gamma_b(2Q)}{\Gamma_b(Q)^3}\frac{\prod \Gamma_b\left(  \tfrac{Q}{2} \pm i P_i \pm i P_j \pm i P_k\right)}{\prod_{a \in \{i,j,k\}}  \Gamma_b \left( Q + 2i P_a \right) \Gamma_b \left( Q - 2i P_a \right)}\,.
\end{equation}
This formula is written in Liouville notation, namely $c=1+6Q^2$ and $h=Q^2/4+P^2$. 
\end{itemize}
This is how the ensemble is defined to leading order at large $c$. We now comment on the differences between an ensemble of this type and an ensemble of approximate CFTs as we have defined in the previous section.

\subsection*{Differences with an ensemble of approximate CFTs}

There are several differences between this ensemble and the ensemble of approximate CFTs as we have defined in this paper. We spell out the most important ones
\begin{itemize}
\item Firstly, as we have already mentioned, an ensemble of approximate CFTs necessarily contains non-trivial non-Gaussian moments, as these are generically implied by approximately imposing certain CFT constraints, such as crossing of higher-point functions. 

\item Secondly, this ensemble is not an ensemble over approximate CFTs in the sense we have defined it. One can see that there is a finite probability to violate certain CFT constraints like modular invariance at genus-two by arbitrarily large amounts. Note that the average of these constraints is still preserved in the ensemble, as is discussed in \cite{Chandra:2022bqq}. This occurs because the deviations from the average value of the constraints come with both signs, and they average to zero. However, there is still a finite probability density to violate the constraints by arbitrarily large amounts. The goal of the rest of the section will be to add non-Gaussianities to the ensemble in order to reduce violations of crossing.
\item There is also a difference at the level of spin. In an ensemble over approximate CFTs, spin remains quantized since it is quantized in every member of the ensemble, whereas in this ensemble it is not.
\item Finally, there is a more subtle difference if conical defects are included. For ensembles of approximate CFTs, while not strictly necessary, it is most convenient to completely fix the data below $\Delta_{\max}$. This is true both of the scaling dimensions \textit{and} the OPE coefficients. Here, for the conical defect operators, the scaling dimensions are fixed but their OPE coefficients are randomized.\footnote{Note that these small fluctuations of the OPE coefficients of conical defect operators are required by the existence of certain wormholes, where conical defects run around non-contractible cycles.} We could easily accommodate this effect with a slightly different choice of probability distribution.
\end{itemize}

Keeping these differences in mind, we are now ready to study the variance of the crossing equation.

\subsubsection{Non-vanishing variance of crossing}
We will now study the variance of the crossing equation. The main result of this section is that the Gaussian ensemble has a non-vanishing (and large) variance for the crossing equation. As we will show, adding a non-Gaussianity to the distribution of OPE coefficients can fix this, drastically reducing the variance of the crossing equation.

\subsubsection*{Pure gravity without conical defects}

A first possibility is to consider gravity without conical defects. In this case, one cannot study the crossing equation of local correlation functions, since there are no local operators which we can use as external operators in correlation functions ($\Delta_{\max}$ is the black hole threshold so all primary operators are above $\Delta_{\max}$). The observables are thus partition functions on compact Riemann surfaces with no operator insertions. Since the density of states is assumed to be classical in the ensemble and the torus partition function only depends on the density of states, powers of the torus partition function factorize into one-point functions and there is no non-trivial variance to compute.

The simplest observable which has a non-zero variance is thus the genus-two partition function, which expanded in the sunset channel reads
\be \label{genus2part}
Z_{\textrm{sunset}}(\Omega,\bar\Omega)\approx \sum_{i,j,k} |C_{ijk}|^2 q_1^{\Delta_i }q_2^{\Delta_j}  q_3^{\Delta_k} \,,
\ee
where we are being schematic about both the choice of coordinates on the moduli space of genus-two surfaces (the relation between the $q_i$ here and the period matrix $\Omega$), and the contribution of conformal blocks. The important feature here is that the partition function is quadratic in the OPE coefficients. The same partition function can also be expanded in a cross-channel, called the dumbbell channel
\be \label{genus2partdumb}
Z_{\textrm{dumbbell}}(\Omega',\bar\Omega')\approx \sum_{l,m,n} C_{llm}C_{mnn} (q'_1)^{\Delta_l }(q'_2)^{\Delta_m}  (q_3')^{\Delta_n} \,,
\ee
for different moduli $\Omega'$. Note that this quantity still depends quadratically on the OPE coefficients, but the contraction is different. This is very similar to a dual channel in four-point crossing (and in fact follows from it). Modular transformations can map \eqref{genus2part} to \eqref{genus2partdumb} upon appropriate mapping of the moduli.

We now study the variance of the genus-two modular crossing equation, given by
\be \label{genus2var}
\Delta Z_{g=2}^2=\overline{(Z_{\textrm{sunset}}(\Omega,\bar{\Omega})-Z_{\textrm{dumbbell}}(\Omega_\gamma,\bar{\Omega}_\gamma))^2} -\Big(\overline{Z_{\textrm{sunset}}(\Omega,\bar{\Omega})-Z_{\textrm{dumbbell}}(\Omega_\gamma,\bar{\Omega}_\gamma)}\Big)^2  \,,
\ee
where $\gamma$ represents an element of the modular group of genus-two surfaces $SP(4,\mathbb{Z})$. There are various choices of modular transformations at genus-two, and one could study the variance for any of them, but we will chose one that maps short cycles in the sunset channel to long cycles in the dumbbell channel.  The second term of \rref{genus2var} vanishes in the large$-c$ limit (this means that the average of the modular crossing equation vanishes, see \cite{Chandra:2022bqq}), but the first one does not. Note that each genus-two partition function involves two OPE coefficients, so the square involves an ensemble average of four OPE coefficients. In the Gaussian ensemble, following \cite{Chandra:2022bqq}, one finds at leading order\footnote{At leading order, only the Gaussian contractions in the diagonal channels (sunset-sunset and dumbbell-dumbbell) give large contributions. In the off-diagonal channels (sunset-dumbbell), there also non-trivial contribution but they are exponentially suppressed compared to the diagonal contributions, resulting from setting many more indices of the OPE coefficients to be the same. We thus chose to not write them here.}
\be \label{nonzerovarg2}
\Delta Z_{g=2}^2 \approx 2Z_{\textrm{sunset}}^L(\Omega,-\Omega)Z_{\textrm{sunset}}^L(-\bar{\Omega},\bar{\Omega}) \,,
\ee
where $Z_{g=2}^L$ is the genus-two partition function computed in Liouville theory. To obtain this expression, we have used the fact that
\be\label{eq.modinv}
\overline{Z_{\textrm{sunset}}(\Omega,\bar{\Omega})Z_{\textrm{sunset}}(\Omega,\bar{\Omega})}\approx\overline{Z_{\textrm{dumbbell}}(\Omega_\gamma,\bar{\Omega}_\gamma)Z_{\textrm{dumbbell}}(\Omega_\gamma,\bar{\Omega}_\gamma)}\approx Z_{\textrm{sunset}}^L(\Omega,-\Omega) Z_{\textrm{sunset}}^L(-\bar{\Omega},\bar{\Omega}) \,,
\ee
which comes from modular invariance under $SP(4,\mathbb{Z})$ transformations acting \emph{simultaneously} on both partition functions. We see that the variance of the genus-two modular crossing equation \rref{nonzerovarg2} is non-zero in the Gaussian ensemble, and is quite large for general values of the moduli. This is fixed by adding a non-Gaussianity to the distribution of OPE coefficients as we demonstrate in the following section.
To simplify our analysis, we will work with an ensemble that includes operators whose conformal dimension scales with $c$ but are below the black hole threshold. Such operators are dual to conical defects in the gravitational theory \cite{Chandra:2022bqq}. The advantage is that we can compute local correlation functions (in particular four-point functions) of these operators creating conical defects, which is slightly simpler than genus-two partition functions.

The connection between the correlation function of conical defect operators and genus-two partition functions can be understood following \cite{Chandra:2022bqq}. Consider a four-point function of conical defect operators and let us start increasing the weights of the conical defect operators until we reach the black hole threshold. Once the operators reach the black hole threshold, one should really sum over these external operators as we can not single out an individual black hole microstate. The (weighted) sum of such four-point functions is the same as a genus-two partition function.
Therefore, the calculations we present in the following section can also be adapted to fix the variance of the genus-two partition function, provided we perform a weighted sum over the external operators. The weight of each term becomes a Boltzman factor $q^\Delta$ where $q$ now refers to a modulus of the Riemann surface. The upshot is that individual heavy operators can violate four-point crossing, but only for a few outliers among the many heavy operators. On average (over the heavy operators), crossing must be obeyed since when averaged over the many heavy operators, it becomes modular crossing at genus-two.

\subsubsection*{Four-point crossing of conical defects}

We are now ready to study the variance of the crossing equation for a four-point function of conical defect operators. We will first do so in the purely Gaussian ensemble, which will motivate the introduction of specific non-Gaussian moments that we will turn to in Section \ref{sec.IntroduceNonGaussianMoment} below. We will chose the following correlation function
\be \label{4ptfctdef}
G_{1122}(x)\equiv \langle O_1(0)O_2(x)O_2(1)O_1(\infty)\rangle \,,
\ee
namely a correlation function with two pairs of distinct conical defect operators. Picking two-distinct pairs makes the distinction between the Gaussian contributions and non-Gaussian contributions more conspicuous.\footnote{One could also have picked four distinct operators, but in that case the mean of the correlation function vanishes identically. Here, the mean of the correlator is non-zero and can be seen to satisfy crossing.} We will use subscripts $\langle\cdot\rangle_{{\rm s/t}}$ to distinguish the correlation functions expanded in the s/t-channel conformal blocks respectively,
\begin{align}
\label{eq.sblock}	G_{1122}(x)|_{\rm s} 
								&= \sum_k |C_{12k}|^2 |\mathcal{F}_{1221}(O_k|x)|^2 \\
\label{eq.tblock}	G_{1122}(x)|_{\rm t} 
								&= \sum_{k'} C_{11k'}C_{22k'} |\mathcal{F}_{1122}(O_{k'}|1-x)|^2~.
\end{align}
In the above expression, $\mathcal{F}$ is a Virasoro block corresponding to the Verma modules of the operators $O_k, O_{k'}$ that are exchanged in the respective channels.

Associativity of the OPE expansion within a correlation function implies that the four-point function is the same whether computed using \eqref{eq.sblock} or \eqref{eq.tblock}. We can see that this is true on average over the ensemble. In the $t$-channel, we have
\begin{align} \label{tchannelaverage}
\overline{G_{1122}(x)|_{\rm t} } 
								&= \sum_{k'} \overline{C_{11k'}C_{22k'} }|\mathcal{F}_{1122}(h_{k'},1-x)|^2=|\mathcal{F}_{1122}(\mathbb{I},1-x)|^2  \,,
\end{align}
which is a single Virasoro block, that of the identity operator. Recall that by definition, products of OPE coefficients with three non-identity indices only have non-vanishing expectation values if the sets of indices agree, as in \eqref{Gaussianmoment}. In the $s$-channel, we have
\begin{align}	\label{schannelaverage}
\overline{G_{1122}(x)|_{\rm s} } 
								&= \sum_k \overline{|C_{12k}|^2} |\mathcal{F}_{1221}(h_k,x)|^2\approx\left|\int_{\frac{c-1}{24}}^{\infty} dh  \rho_0(h)C_0(h_1,h_2,h) \mathcal{F}_{1221}(h,x)  \right|^2 \,.
\end{align}
In this channel, the entire family of heavy operators contributes with OPE coefficients given by the $C_0$ formula. Notice that the derivation of $C_0$ precisely follows from crossing symmetry where one keeps only the identity in the cross-channel, which is exactly what we have in \rref{tchannelaverage}. We thus find

\begin{equation} \label{meanof4ptcrossing}
	\overline{G_{1122}(x)|_{\rm s}  - G_{1122}(x)|_{\rm t} } \approx 0~.
\end{equation}

Note the ``$\approx$'' sign in the crossing equation. There are exponentially small corrections that we have not taken into account here. As explained in \cite{Chandra:2022bqq}, in the $s$-channel, the operators $O_1$ and $O_2$ can fuse to any conical defect operator, which is not included in the integral of \rref{schannelaverage}. While we could easily include them in the $s$-channel, they would produce a missmatch in the $t$-channel where the nature of the ensemble is such that literally only the identity contributes. To produce a fully crossing symmetric average, one would need to correct the ensemble of the $C_{ijk}.$\footnote{Note that in \cite{Chandra:2022bqq}, they call this a non-Gaussianity. However, this missmatch could still be fixed by a purely Gaussian ensemble, but the OPE coefficients would no longer be independent Gaussian random variables, but rather they would be correlated such that $\overline{C_{11j}C_{22j}}\neq 0$.} We emphasize here that these are exponentially small corrections, which we will neglect. Next we turn to the variance of the crossing equation.

The quadratic moment of the crossing equation reads
\begin{align}
	\Delta G_{1122}(x,x')=\overline{\left(G_{1122}(x)|_{\rm s}  - G_{1122}(x)|_{\rm t} \right)\left(G_{1122}(x')|_{\rm s}  - G_{1122}(x')|_{\rm t} \right)}~ \,.
\end{align}
The diagonal expression in cross-ratios (i.e. $x=x'$) is simply the variance of the crossing equation. We will now show that this variance is non-vanishing at leading order in the large $c$ expansion, and we will need to add the relevant non-Gaussian correction to the distribution of OPE coefficients in order to fix this problem. The expansion of the above observable can be understood as a sum over products of four point functions defined in different copies of the theory,
\begin{align}\label{eq.variance}
	\Delta G(x,x') &= \overline{G_{1122}(x)|_{\rm s}G_{1122}(x')|_{\rm s}} + \overline{G_{1122}(x)|_{\rm t}G_{1122}(x')|_{\rm t}} \nonumber\\
&\hspace{-15pt}- \overline{G_{1122}(x)|_{\rm s}G_{1122}(x')|_{\rm t}} - \overline{G_{1122}(x)|_{\rm t}G_{1122}(x')|_{\rm s}}~.
\end{align}
We will refer to each term in this sum as a \emph{two-sided observable}, in analogy with their gravitational interpretation in terms of two-sided Euclidean wormhole geometries \cite{Chandra:2022bqq}. A given term can be expanded in the following way
\begin{align}
	\label{eq.schem-avg}	\overline{G_{1122}(x)|_{\mathfrak c_1}G_{1122}(x')|_{\mathfrak c_2}} &= \int\limits_{}^{}\!\!{\rm d}\!\left[ O_6\right]{\rm d}\!\left[ O_{6'} \right] \rho([O_6])\rho([O_{6'}]) ~\overline{C_{\mathfrak c_1}^2 C_{\mathfrak c_2}^2}~ |\mathcal{F}_{\mathfrak c_1}(O_{6}|x_{\mathfrak c_1})|^2|\mathcal{F}_{\mathfrak c_2}(O_{6'}|x_{\mathfrak c_2}')|^2~.
\end{align}
Here, $\mathfrak c_{1,2}$ denotes the channel in which we expand the correlation function, which affects both the associated cross-ratios and blocks. 
We have also used the notation $\rm{d}[O]$ to denote the measure for an operator, and $\rho([O])$ as a shorthand to denote the density of such operators. We also see the appearance of an ensemble average over four OPE coefficients. There are various contributions to the ensemble average
\be
\overline{C_{\mathfrak c_1}^2 C_{\mathfrak c_2}^2} \,.
\ee
Since the ensemble is (for now) Gaussian, we simply need to keep track of all the Wick contractions of the OPE coefficients. The simplest contributions are the disconnected contributions. These give 
\begin{align}\label{eq.variancezero}
	\Delta G(x,x') &= \overline{G_{1122}(x)|_{\rm s}}\ \overline{G_{1122}(x')|_{\rm s}} + \overline{G_{1122}(x)|_{\rm t}}\ \overline{G_{1122}(x')|_{\rm t}} \nonumber\\
														    &\hspace{-15pt}- \overline{G_{1122}(x)|_{\rm s}}\ \overline{G_{1122}(x')|_{\rm t}} - \overline{G_{1122}(x)|_{\rm t}}\ \overline{G_{1122}(x')|_{\rm s}}~. \notag \\
                &\approx 0 \,.      
\end{align}
Here, we have used \rref{meanof4ptcrossing}, namely that the mean of the crossing equation is approximately zero. Therefore, we only need to study connected contractions between OPE coefficients.

There are essentially two types of two-sided correlators: $s-s,t-t$ or $s-t,t-s$ (namely whether each individual correlator is expanded in the same channel, or in different channels). For $s-s$ or $t-t$ correlators, we find the following connected Wick contractions\footnote{In the first line below, we used the fact that $C_{ijk}=C_{ikj}^*=(-1)^{(J_i+J_j+J_k)}C_{ikj}$, where $J_i$ is the spin of the operators.}
\bea \label{6eq6pr}
\int\limits_{}^{}\!\!{\rm d}\!\left[ O_6\right]{\rm d}\!\left[ O_{6'} \right] \rho([O_6])\rho([O_{6'}])\overline{C_{ 116} C_{ 226} C_{ 116'} C_{ 226'}} &=& \int\limits_{}^{}\!\!{\rm d}\!\left[ O_6\right]\rho([O_6]))\overline{|C_{ 116}|^2 }\ \overline{ |C_{ 226}|^2}\\
\int\limits_{}^{}\!\!{\rm d}\!\left[ O_6\right]{\rm d}\!\left[ O_{6'} \right] \rho([O_6])\rho([O_{6'}])\overline{|C_{ 126} |^2| C_{ 126'} |^2} &=& \int\limits_{}^{}\!\!{\rm d}\!\left[ O_6\right]\rho([O_6]))\overline{|C_{ 126}|^2 }\ \overline{ |C_{ 126}|^2}\,.
\eea
The connected contribution in the $s-s$ and $t-t$ channels give a Liouville four-point function \cite{Chandra:2022bqq}, albeit with un-conventional dependence on cross ratios,
\be \label{ss}
\textrm{s-s channel}|_{\textrm{con}}=G^L_{1221}(x,x')G^L_{1221}(\bar{x},\bar{x}') \,.
\ee
In the $t-t$ channel, we have
\be \label{tt}
\textrm{t-t channel}|_{\textrm{con}}=G^L_{1122}(1-x,1-x')G^L_{1122}(1-\bar{x},1-\bar{x}') \,.
\ee
In the $s-t$ and $t-s$ channels, we get no contributions as
\be \label{stzero}
\overline{C_{ 126}C_{ 116'} }= \overline{C_{ 126}C_{ 226'} }=0 \quad \forall  \ 6,6' \,.
\ee
Note that that the $s-s$ and $t-t$ contributions are actually the same, since the Liouville correlator is crossing invariant, and the difference between the $s-s$ and $t-t$ contributions is a simultaneous transformation on the cross ratios $x$ and $x'$. This means that in total, we find
\be \label{nonzerovar4pt}
\Delta G (x,x') = 2 G^L_{1221}(x,x')G^L_{1221}(\bar{x},\bar{x}') \,.
\ee
Setting $x=x'$, we find the variance of the crossing equation
\be
\Delta G (x,x) = 2 G^L_{1221}(x,x)G^L_{1221}(\bar{x},\bar{x}) \,.
\ee
This follows closely what we found for modular crossing at genus-two in \rref{nonzerovarg2}. 

We will now modify the model by introducing a non-trivial quartic moment for the OPE coefficients. As we will see, this will reduce the variance of the crossing equation and cancel the non-zero term we found in the right-hand side of \rref{nonzerovar4pt}.

\subsection{Fixing the variance by introducing non-Gaussianities}\label{sec.IntroduceNonGaussianMoment}

We now consider a modified model, where on top of the Gaussian contribution to the moments of OPE coefficients \rref{Gaussianmoment}, we add a non-trivial quartic moment. The quartic moment is of the following form 
\be \label{eq.4cnonGauss} 
\overline{C_{ijk} C_{iml} C_{njl} C_{nmk}} \Big|_c= \begin{Bmatrix}
        \mathcal O_k&\mathcal O_j&\mathcal O_i \\
        \mathcal O_l&\mathcal O_m&\mathcal O_n
    \end{Bmatrix}= \left|\mathbb F_{kl}\begin{bmatrix}
        n&j\\
        m&i
    \end{bmatrix} C_0(h_i,h_j,h_k) C_0(h_k,h_n,h_m)\rho_0(h_l)^{-1} \right|^2 
     \,,
\ee
where $c$ stands for connected (i.e. subtracting the Gaussian contractions),  $\begin{Bmatrix}
        \mathcal O_k&\mathcal O_j&\mathcal O_i \\
        \mathcal O_l&\mathcal O_m&\mathcal O_n
    \end{Bmatrix}$ is the Virasoro 6-j symbol\footnote{Our conventions for the 6j symbol follow the original conventions of Wigner.} and $\mathbb{F}$ is the Virasoro crossing kernel. The crossing kernel implements the change of basis from $s$ to $t$ channel Virasoso blocks, and is defined as 
\be \label{crossingkerneldef}
\mathcal{F}_{ijmn}\left( O_{k}|x \right)= \int\limits_{}^{}\!\!{\rm{d}}[O_{l}]  \,\mathbb F_{kl}\begin{bmatrix}
        n&j\\
        m&i
    \end{bmatrix}  \mathcal{F}_{imjn}\left( O_{l}|1-x \right) \,,
\ee
   The crossing kernel is known in closed form \cite{Ponsot:1999uf,Ponsot:2000mt} and can be studied explicitly in various limits (see \cite{Collier:2018exn,Collier:2019weq} for detailed reviews). It also plays a central role in the derivation of the $C_0$ formula. From \eqref{eq.4cnonGauss}, we see the close connection between the crossing kernel and the 6j symbol.

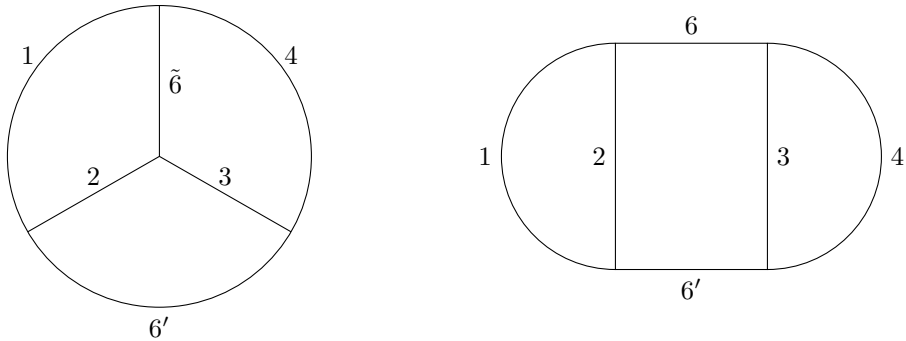
\begin{figure}[tpb]
    \centering
    \begin{tikzpicture}
        \draw (-4,1) circle[radius = 2];
        \draw (-4,3) -- (-4,1) node[midway, right]{$\tilde6$} -- (-2.268,0) node[midway, above]{3};
        \draw (-4,1) -- (-5.732,0) node[midway, above]{2};
        \node[below] at (-4,-1) {$6'$};
        \node[above] at (-2.268,2.1) {4};
        \node[above] at (-5.732,2.1) {1};
        \draw (2,-0.5)rectangle (4,2.5) ;
        \draw (2,-0.5) arc (270:90:1.5);
        \draw (4,-0.5) arc (-90:90:1.5);
        \node[left] at (0.5,1) {1};
        \node[right] at (5.5,1) {4};
        \node[above] at (3,2.5) {$6$};
        \node[below] at (3,-0.5) {$6'$};
        \node[left] at (2,1) {2};
        \node[right] at (4,1) {3};
    \end{tikzpicture}
    \caption{A genus-three partition function can be decomposed in several different channels. We show two examples: the skyline channel (left), which is the important non-Gaussianity that we focus on here, and the pillow channel (right). The dominant contribution to the pillow channel comes from a Gaussian contraction which only takes into account the terms with $6=6'$ \cite{Chandra:2022bqq}.}
    \label{fig.pillow-sky}
\end{figure}

The formula \rref{eq.4cnonGauss} is a non-Gaussianity coming from modular invariance at genus-three \cite{Belin:2021ryy}, and represents an OPE index contraction known as the skyline channel (see Fig. \ref{fig.pillow-sky}).\footnote{Here, because we are studying four-point functions of conical defects, we have taken some of the operators (the external ones) to lie below the black hole threshold, and have analytically continued the formula. This is similar to what is done in \cite{Chandra:2022bqq} directly at the level of the $C_0$ formula for OPE coefficients of three conical defects. To study the variance of modular crossing at genus-two, one would use the same non-Gaussianity without the analytic continuation.} Even though this is a genuine non-Gaussianity, we see that it is related to a Gaussian contraction by the addition of the crossing kernel which connects all the indices. All the non-Gaussianities (quartic and higher) are of this type \cite{Belin:2021ryy}, and come from applying iterative crossing moves. This fact will be important in section \ref{sec.tensor}.

We will now show that this single non-Gaussianity is sufficient to restore a vanishing variance of crossing (to leading order in the large $c$ limit). Let us return to the study of the $s-t$ cross terms that vanished in the Gaussian model due to \rref{stzero}. We have 

\begin{align}
    \left.\overline{G_{1122}(x)|_{\rm s}G_{1122}(x')|_{\rm t} }\right|_c &=\int\limits_{}^{}\!\!{\rm{d}}[O_{6'}]{\rm{d}}[O_{\tilde6}] \, \rho([O_{6'}])\, \rho([O_{\tilde6}]) \,\overline{|C_{126'}|^2 C_{11\tilde{6}}C_{22\tilde{6}}}|_c \\
    & \hspace{3cm}\times |\mathcal{F}_{1221} \left( O_{6'}|x \right) |^2|\mathcal{F}_{1122}\left( O_{\tilde 6}|1-x' \right)|^2 \nonumber \\
    &\approx \int\limits_{}^{}\!\!{\rm{d}}[O_{6'}]{\rm{d}}[O_{\tilde6}] \, \rho([O_{6'}]) \,\left|\mathbb F_{6'\tilde{6}}\begin{bmatrix}
        n&j\\
        m&i
    \end{bmatrix} C_0(h_1,h_2,h_{6'})^2  \right|^2  \nonumber \\
       & \hspace{3cm}\times |\mathcal{F}_{1221} \left( O_{6'}|x \right) \mathcal{F}_{1122}\left( O_{\tilde 6}|1-x' \right) |^2 \,.
       \end{align}
       In the above, we have used the fact that $C_{ijk}=C_{jik}^*$. We can now use the defintion of the crossing kernel \eqref{crossingkerneldef} to get rid of the integral over the operator $\tilde{6}$ by transforming the blocks. We find 
       \begin{align} \label{eq.st2}
       \left.\overline{G_{1122}(x)|_{\rm s}G_{1122}(x')|_{\rm t} }\right|_c
&\approx \int\limits_{}^{}\!\!{\rm{d}}[O_{6'}] \, \rho([O_{6'}])\, |C_0(h_1,h_2,h_{6'})^2| \,  |\mathcal{F}_{1221} \left( O_{6'}|x \right) \mathcal{F}_{1221} \left( O_{6'}|x' \right) |^2 \notag \\
&\approx G^L_{1221}(x,x')G^L_{1221}(\bar{x},\bar{x}') \,.
\end{align}
A similar treatment of the $t-s$ channel gives
\begin{align}
    \left.\overline{\langle O_1 O_2 O_2 O_1 \rangle_{\rm t}\langle O_1 O_2 O_2 O_1 \rangle_{\rm s}}\right|_c 
    \label{eq.ts2}  &\approx G^L_{1221}(1-x,1-x')G^L_{1221}(1-\bar{x},1-\bar{x}') ~.
\end{align}
Adding the $s-s$ and $t-t$ channel whose leading order contribution comes from the Gaussian moment, we find
\be \label{zerovar4pt}
\Delta G(x,x') \approx 0 \,,
\ee
Setting $x=x'$, this calculation establishes that the variance of the crossing equation is very small in this model with a quartic non-Gaussianity, and vanishes to leading order in the large $c$ approximation. It would be straightforward to generalize this statement to modular invariance at genus-two, simply by pushing the operators $O_{1,2}$ above the black hole threshold and summing over them. The same non-Gaussianity would thus also enforce that the variance \rref{nonzerovarg2} vanishes.

An interesting interpretation of the derivations above is as follows. One can use the variance of crossing in order to pin down the fourth moment of the statistics of the OPE coefficients. In principle, in this way, one can determine the higher moments in an analogous fashion, thereby building up the ensemble from its moments. Interestingly, the fourth moment one constructs in this way exactly coincides with the non-Gaussianity obtained from considering modular invariance of the genus-three partition function. We similarly expect higher moments to match results obtained from modular invariance at higher genus. We note, however, that this procedure is not taking into account various types of large$-c$ non-perturbative corrections, which we discuss in the following section.

\subsubsection{Subleading corrections to the variance of the crossing equation}\label{sec.subleadingcorr}

We have just seen that a quartic non-Gaussianity can produce a large contribution in the $s-t$ (and $t-s$) channel, cancelling the Gaussian contributions in the diagonal channels. We established this statement at the leading non-vanishing order in the large $c$ limit (up to corrections that are exponentially suppressed in $c$), and we now discuss corrections.
There are essentially two types of corrections that can appear at subleading orders. A first source of correction comes from the permutation of indices in the Gaussian and quartic moments \eqref{Gaussianmoment} and \eqref{eq.4cnonGauss}. In our calculations, we have always included the leading arrangement of indices that maximally ``click'', but subleading contributions also appear where more indices are set to be the same. A second type of correction would come from studying the exchange of sub-threshold states, which require introducing a non-diagonal moment $\overline{C_{11k}C_{22k}}$, see \cite{Chandra:2022bqq}. 

In the end, we cannot currently tell if the different corrections cancel in the variance or not, and whether they give bounded contributions as a function of the cross-ratio. It would be interesting if these corrections exactly cancelled. Note that we have considered an ensemble without correlations in the spectrum, but we now discuss the type of exponential corrections that would arise if spectral correlations were included.

\subsubsection*{Adding spectral correlations}

So far we have not considered spectral correlations in the variance of crossing, and now discuss their effect if they were included in the ensemble. If spectral correlations are included in the square of correlation functions, we would obtain from \eqref{eq.schem-avg}
\begin{align}
	\overline{\rho([O_{6'}]) \rho([O_6]) C^2_{\mathfrak c_1} C^2_{\mathfrak c_2}}~.
\end{align}
We have only considered the contributions of the following kind in our analysis:
\begin{align}
    \overline{\rho([O_{6'}])} ~ \overline{\rho([O_6])}~& \overline{C^2_{\mathfrak c_1}}~ \overline{C^2_{\mathfrak c_2}}~,\\
\label{eq.leadingcontract}    \overline{\rho([O_{6'}])}~ \overline{\rho([O_6])}~& \overline{C_{\mathfrak c_1} C_{\mathfrak c_2}}~ \overline{C_{\mathfrak c_1} C_{\mathfrak c_2}}~.
\end{align}
The first line above gives the disconnected contribution (that we are not so interested in), while the second line computes the leading contribution to the connected piece. We will now study a different connected contribution that gives a (small) correction to this leading connected contribution.
Note that as proposed in \cite{Belin:2021ibv}, the statistics of the OPE coefficients need not be independent of the statistics of the density of states. However, treating them independently is correct at this order. We thus wish to consider
\begin{equation}
	\overline{\rho([O_{6'}]) \rho([O_6])} ~\overline{C^2_{\mathfrak c_1}} ~ \overline{C^2_{\mathfrak c_2}}~.
\end{equation}
This time, the connected nature of the contribution between the two sides comes from the density of states and not the ``contractions'' between OPE coefficients. We now show that the contribution of this term is suppressed compared to the value of the connected part that we have computed in the previous section.
In the computation of the variance of the crossing equation, it corresponds to terms of the type
\begin{align}
    \left. \overline{\langle O_1 O_2 O_2 O_1 \rangle_{\rm s}\langle O_1 O_2 O_2 O_1 \rangle_{\rm s}} \right|_{c} &= \int\limits_{}^{}\!\! {\rm d}\! \left[O_6\right]{\rm d} [O_{6'}]\, \left. \overline{\rho\!\left( \left[O_6\right] \right)\rho\!\left( \left[O_{6'}\right] \right)}\right|_{c} \, \overline{|C_{126'}|^2}\, \overline{|C_{126'}|^2}  \nonumber \\[-10pt]
\label{eq.ss.den.conn}   & \hspace{5cm} \times |\mathcal{F}_{126}(O_{6}|x)|^2|\mathcal{F}_{126'}(O_{6'}|x)|^2~.
\end{align}
In the ergodic limit, assuming that correlations between the density of states are given by the sine kernel \cite{Altland:2020ccq}, we would find
\begin{equation}
	\frac{\overline{\rho([O_{6'}]) \rho([O_6])}|_c}{\overline \rho~ \overline \rho} \approx 1 - \frac{\sin^2s}{s^2} \sim e^{-2S}~.
\end{equation}
Here, $s$ is the energy separation $\Delta_{6'} - \Delta_6$ measured in units of the mean level spacing of the spectrum, i.e. $e^{-S}=e^{-2\pi \sqrt{\frac{c}{3}\Delta}}$. This has been checked explicitly in low-dimensional examples of holography, \cite{Cotler:2020ugk, Cotler:2020lxj, Cotler:2022rud}. We see that connected spectral correlations suppress the answer by $e^{-2S}$, whereas the connected contribution of OPE coefficients only suppressed them by a factor of $e^{-S}$ (coming from setting $O_6=O_{6'}$ in \eqref{6eq6pr}). Thus, would they be included in the ensemble, connected spectral correlations only produce exponentially small corrections to connected two-sided observables.

\subsubsection{Comments on the gravitational dual}
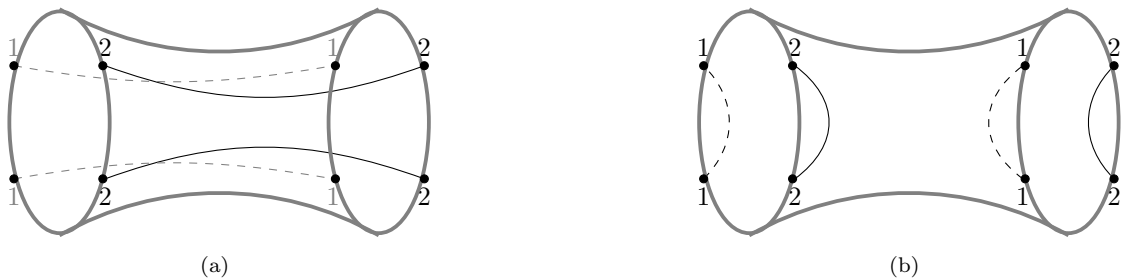
\begin{figure}[tb]
    \centering
    \subfigure[]
    {
      \begin{tikzpicture}[scale=0.3]
        \draw (-5,2.5) node[above]{2} to[out=-20,in=200] (9,2.5) node[above]{2};
        \draw (-5,-2.5) node[below]{2} to[out=20,in=160] (9,-2.5) node[below]{2};
        \draw[dashed, gray] (-9,2.5) node[above]{1} to[out=-10,in=190] (5,2.5) node[above]{1};
        \draw[dashed, gray] (-9,-2.5) node[below]{1} to[out=10,in=170] (5,-2.5) node[below]{1};
        \draw[line width = 0.5mm, gray] (-7,-5) arc(120:60:14);
        \draw[line width = 0.5mm, gray] (-7,5) arc(240:300:14);
        \draw[line width = 0.5mm, gray] (7,0) ellipse (2.2cm and 4.9cm);
        \draw[line width = 0.5mm, gray] (-7,0) ellipse (2.2cm and 4.9cm);
        \filldraw (-5.1,-2.5) circle (5pt);
        \filldraw (-5.1,2.5) circle (5pt);
        \filldraw (-9,-2.5) circle (5pt);
        \filldraw (-9,2.5) circle (5pt);
        \filldraw (5.1,-2.5) circle (5pt);
        \filldraw (5.1,2.5) circle (5pt);
        \filldraw (9,-2.5) circle (5pt);
        \filldraw (9,2.5) circle (5pt);
    \end{tikzpicture}
    \label{fig.wormhole.ope.conn}
    }
    \hfill
    \subfigure[]
    {
     \begin{tikzpicture}[scale=0.3]
        \draw (-5,-2.5) node[below]{2} .. controls (-3,-1) and (-3,1) .. (-5,2.5) node[above]{2};
        \draw[dashed] (-9,-2.5) node[below]{1} .. controls (-7.5,-1) and (-7.5,1) .. (-9,2.5) node[above]{1};
        \draw[dashed] (5,-2.5) node[below]{1} .. controls (3,-1) and (3,1) .. (5,2.5) node[above]{1};
        \draw (9,-2.5) node[below]{2} .. controls (7.5,-1) and (7.5,1) .. (9,2.5) node[above]{2};
        \draw[line width = 0.5mm, gray] (-7,-5) arc(120:60:14);
        \draw[line width = 0.5mm, gray] (-7,5) arc(240:300:14);
        \draw[line width = 0.5mm, gray] (7,0) ellipse (2.2cm and 4.9cm);
        \draw[line width = 0.5mm, gray] (-7,0) ellipse (2.2cm and 4.9cm);
        \filldraw (-5.1,-2.5) circle (5pt);
        \filldraw (-5.1,2.5) circle (5pt);
        \filldraw (-9,-2.5) circle (5pt);
        \filldraw (-9,2.5) circle (5pt);
        \filldraw (5.1,-2.5) circle (5pt);
        \filldraw (5.1,2.5) circle (5pt);
        \filldraw (9,-2.5) circle (5pt);
        \filldraw (9,2.5) circle (5pt);
    \end{tikzpicture}
    \label{fig.wormhole.den.conn}
    }
    \caption{(a) A bulk geometry that computes the connected contribution of a square of four-point functions. The action of this geometry matches \rref{ss}. (b) Diagrammatic representation of a putative bulk geometry that would contribute to the connected correlations of spectral densities, as estimated in \eqref{eq.ss.den.conn}. The OPE coefficients on the two sides are \emph{not} correlated. There is no known saddle that captures this correlation, and this geometry may be off-shell.}
\end{figure}

So far, we have performed all the computations in the ``CFT" ensemble. However, there are gravitational counterparts to many of these calculations, in the form of Euclidean wormholes. As we already mentioned, this was the reasoning behind the terminology \textit{two-sided observable}. As explained in \cite{Chandra:2022bqq}, the connected contribution \eqref{ss} matches the on-shell action and 1-loop determinant of a two-boundary Euclidean wormhole, with four conical defects propagating through the bulk, see Fig. \ref{fig.wormhole.ope.conn}. A natural question to ask is whether contributions coming from non-Gaussianities give rise to new bulk geometries.

In general, the answer to this question is yes. There will be new geometries that do not exist in the Gaussian ensemble. One example of such a geometry was recently discussed in \cite{Collier:2023fwi}: a four-boundary wormhole with 4 spherical boundaries, each having 3 conical defect insertions. One expects many other geometries to appear as well, involving more boundaries or higher topologies. For the particular example we discussed above, namely the product of two four-point functions with pairwise identical operators, the non-Gaussianities do not produce any new geometries. This can be seen from the fact that the non-Gaussianity does not produce any new contribution to the $s-s$ channel. It would be interesting to see if new geometries resulting from a non-Gaussianity appear for other squares of observables, like the square of the genus-two partition function. We expect this to be the case, but we also expect their contribution to precisely cancel in the variance.

It is also interesting to ask whether correlation functions of spectral densities can be matched to bulk geometries. For the product of torus partition functions, these correlations are not captured by saddles, but rather by off-shell geometries \cite{Saad:2018bqo,Cotler:2020ugk,Cotler:2020lxj}. We also expect that off-shell geometries would capture such correlations in other observables, as we have estimated in \eqref{eq.ss.den.conn}. We represent a putative geometry of this type in Fig. \ref{fig.wormhole.den.conn}.

\section{A tensor model for 3D gravity: Virasoro simplicial gravity}\label{sec.tensor}
In this section we will describe a model of gravity, built on our notion of approximate CFT. We consider the particular case that describes pure gravity in 3-dimensions (we will chose to not add conical defects, but it would be a straightforward generalization). Thus the approximate CFTs that we consider have a large gap and all the states lie above the blackhole threshold $\Delta>\frac {c-1}{12}$, with the vacuum state being the only unique sub-threshold state. We describe a candidate dual ensemble of 3D gravity, as an explicit ensemble of the CFT data of heavy states, with a probability measure that weights the data by a factor that decays rapidly away from solutions to certain crossing relations. In particular we will consider modular crossing (the modular S-transform) for torus one-point functions, as well as standard crossing of sphere four-point functions. In each case, this leads to a measure of the form 
\begin{equation}
e^{- a \sum_I |F_I|^2} \,,
\end{equation}
where $F_I = 0$ are crossing relations, considered block by block (see below), and the multi-index $I$ labels sets of heavy external operators.\footnote{Notice that crossing symmetry for individual heavy operators was not part of our original requirements for an approximate CFTs. Smeared versions of crossing symmetry of heavy operators are however needed in order to establish modular invariance on higher genus surfaces. Requiring approximate crossing for individual heavy operators might therefore be too strong a condition from the approximate CFT perspective, and we will come back to the question of smearing momentarily.} The details of the constraints, as well as the composition of the multi-index depend on the details of the particular crossing constraint under consideration and will be spelled out shortly. The resulting potential is a sum of squares, thus the limit $a \rightarrow \infty$ imposes the constraints exactly. The outcome of this procedure depends on whether pure gravity solutions to the CFT bootstrap exist or not, which is currently an open problem.\footnote{Answering this question is one of the goals of the modular bootstrap program initiated in \cite{Hellerman:2009bu}, see \cite{Maloney:2007ud,Keller:2014xba,Hartman:2019pcd,Afkhami-Jeddi:2020hde,Benjamin:2020mfz,Maxfield:2020ale,DiUbaldo:2023hkc} for its connection to pure 3d gravity. Our matrix tensor model may be a fresh tool to revisit this important question from a different technical angle.}  If such solutions exist, the $a\to\infty$ limit would lead to an ensemble of exact CFTs, which is presumably a sum of delta functions as a distribution over the CFT data. In accordance with our definition of approximate CFTs, for now we take $a$ to be finite but large. This parameter controls the average size of the violation of the crossing constraints in typical members of the ensemble.

We thus construct an ensemble of random matrices $\Delta_{ij}$, and three-tensors $C_{ijk}$, which corresponds to an approximate  CFT truncated to the first $K$ operators, where $\Delta$ is the dilatation operator, acting as a matrix on the vector space of primaries and $C$  is the rank 3 symmetric tensor of structure constants. The definitions above - to be made more precise in the rest of this section - have introduced two parameters, namely the truncation parameter $K$, and the constraint parameter $a$. The first of these renders the matrices and tensors finite, while the second expresses the amount by which the resulting model may fluctuate around the exact solutions of the constraint equations. The second issue was already explored in section \ref{sec.AveragingApproxCFT}, while the truncation $K$ is a (necessary) new feature of the tensor model. A continuum limit of the model will have to be defined via a double-scaling type procedure in which both regulators are taken to infinity simultaneously, leaving an appropriate ratio finite. We do not perform this double scaling limit, and instead leave the resulting continuum limit for future work. It will be important to define and perform this limiting procedure carefully, most likely in analogy to the double-scaling performed in \cite{Jafferis:2022uhu,Jafferis:2022wez}.

In fact, since we are interested in the case of two-dimensional CFTs, it is easiest to consider left and right weights $(h_i, \bar h_i)$ as independent random variables, which are separately integrated over, as two mutually commuting matrices. In our truncated model of CFT data, $(h_i, \bar h_i)$ are the eigenvalues of the {\it finite} matrix representations of $L_0$ and $\bar L_0$. Note also that we are integrating over the dimensions as well as a continuous spin variable, as is appropriate in the asymptotic regime of \cite{cardyformula, Hartman:2014oaa}. Integrality of spin across all regimes can be exactly imposed by requiring T-modular invariance of the potential, and we return to this point below. We start by integrating with the flat measure on the components, however the full measure will include a potential term, determined by imposing (approximate) crossing constraints on the matrix and tensor variables. We thus propose a random matrix-tensor model with action given schematically by

\begin{equation}
{\cal Z } = \int D[L_0,\bar L_0]\, D [C]\ e^{ - a\, V[L_0, \bar L_0, C] }\,,
\end{equation} 
where the potential $V$ will be determined below. This model is invariant under unitary transformations acting on the vector space of primaries, although it will often be convenient to write our expressions in the eigenbasis of $(L_0,\bar L_0)$, as was done in \cite{Jafferis:2022wez,Jafferis:2022uhu}, for an analogous construction of matrix models for JT gravity with matter. Going to the eigenbasis of $L_0,\bar L_0$, introduces a Vandermonde factor, which in the following we will always assume to be included in the measure. Partition functions of the CFT expressed in terms of the dimensions and structure constants are observables that can be inserted into the ensemble. 
If the theory contained\footnote{This is a statement about putative tensor model for approximate CFTs not necessarily dual to pure gravity.} light operators below $\Delta_{\rm max}$, then in the above integration the OPE coefficients for those light operators could in principle appear in the potential but would not be included as integration variables. This follows from our choice of ensembles of approximate CFTs, where we only average ``heavy data'' subject to constraints due to fixing light data (and symmetries). 

We will see that the potential above automatically includes Gaussian terms for the structure constants, as well as large non-Gaussianities, whose origins are the approximate enforcement of crossing constraints on the data. This is a stronger condition, which ensures that for example the variance of higher-genus crossing can be correctly computed. Moreover, the model is fundamentally non-Gaussian and we expect that diagrammatic calculations in this tensor model will contain new contributions associated to other 3d topologies in the special case of pure 3D gravity. Indeed the model explicitly takes on a form previously considered in the simplicial approach to lattice 3D gravity \cite{regge1961general,Boulatov:1992vp,turaev1992state}, with the quartic non-linearity given by a 6j-symbol, although in our case the underlying group structure is Virasoro and not SU$(2)$. Furthermore, our model adds the matrix degrees of freedom $L_0, \bar L_0$, which have no analogs in \cite{regge1961general,Boulatov:1992vp,turaev1992state}.

 This analogy suggests an identification of the factor $a$ with the discrete lattice regulator.  Despite the different underlying group structure, this relation to lattice gravity merits further study.

\subsubsection*{Preliminary comments on the spectrum and notation}
In this section we summarise the conventions used conventionally in Liouville theory for various parameters of the theory like central charge and weights of the operators. The same parametrisation is used to label the Virasoro inversion kernels of the crossing equations. The central charge of the theory is given by,
\begin{equation}
    \label{eq.centralcharge}
    c = 1 + 6 Q^2 = 1 + 6 \left(b+\frac1b\right)^2~.
\end{equation}
The chiral weights of the operators in terms of the Liouville `momentum' is given by,
\begin{align}
    \label{eq.hhbar}
    h = \frac{Q^2}4 + P^2 = \frac{c-1}{24} + P^2, \qquad
    \bar h = \frac{Q^2}4 + \bar P^2 = \frac{c-1}{24} + \bar P^2~.
\end{align}
Alternatively, the weights can also be written in terms of a different parametrisation,
\begin{align}
    \label{eq.hinalpha}
    h &= \alpha \left( Q - \alpha\right), \qquad \bar h = \bar \alpha \left( Q - \bar \alpha\right).
\end{align}
While the parametrisation in \eqref{eq.hhbar} spans the spectrum of operators with $h\ge \frac{c-1}{24}$, the parametrisation \eqref{eq.hinalpha} spans the operators below the threshold. Note that the operators spanned by $P>0$ are above the blackhole threshold for a holographic CFT. The two sets of parametrisation are related by,
\begin{equation}
    \alpha = \frac Q2-iP, \qquad \bar \alpha  = \frac Q2 - i\bar P~.
\end{equation}
Below, whenever applicable, we integrate $P$ and $\bar P$ over the whole real line $\mathbb{R}$ by symmetry, rather than restricting the integral to run over non-negative reals. For integrals over the weights, we absorb all factors of two that arise from this choice in the integration measures. For orthogonality relations (see \eqref{eq.InversionOrtho} for example), we will only write $\delta^{(2)}(P_1-P_2)$, since it is equivalent to $\frac{1}{2}\delta^{(2)}(P_1-P_2)+\frac{1}{2}\delta^{(2)}(P_1+P_2)$ when integrated against even functions of momentum, which will always be the case for us.
\subsection{Virasoro four-point crossing constraints and simplicial gravity}
Let us now proceed to concretely specifiy the potential $V[L_0,\bar L_0, C]$ outlined above. For this it will be useful to recall the crossing relations for torus one-point functions and sphere four-point functions, in the approach of \cite{Collier:2018exn, Collier:2019weq}.  This technique relies on the inversion properties of the conformal blocks that enter the definitions of these physical observables. As a concrete example, let us consider the sphere four-point crossing, which we write as
\begin{equation}\label{eq.BasicVirasoroCrossing}
\sum_p C_{i_1 i_2p}C_{i_3 i_4 p} {\cal F}(p | z) \overline{{\cal F}}(p | \bar z) = \sum_q C_{i_1 i_4q}C_{i_2 i_3 q} {\cal F}(q | 1-z) \overline{{\cal F}}(q | 1 -\bar z) \,,
\end{equation}
where we have chosen a sum notation for the operators appearing in the OPE, which is meant to indicate a sum over discrete parts and an integral with the appropriate density over the continuum parts. For some purposes it may be more convenient to write the same expression in integral form by introducing the spectral density $\rho(p)$, which reduces to a sum of delta function for the discrete parts. One then writes
\begin{equation}
\int\limits_{\mathbb R\cup\mathds 1} \!\!\frac{d^2P_p}4\, |\rho(p)|^2\, C_{i_1 i_2p}C_{i_3 i_4 p} {\cal F}(p | z) \overline{{\cal F}}(p | \bar z) = \int\limits_{\mathbb R\cup\mathds 1} \!\!\frac{d^2P_q}4\, |\rho(q)|^2\, C_{i_1 i_4q}C_{i_2 i_3 q} {\cal F}(q | 1-z) \overline{{\cal F}}(q | 1 -\bar z) \,,
\end{equation}
Above, ${\cal F}$ are the Virasoro blocks whose functional form depends implicitly on the OPE contraction used on each side of the crossing equation, that is values of the indices $i_1\cdots i_4$. The LHS side is the `s-channel' expansion, while the RHS gives the `t-channel' expansion instead. This equation should be seen as a constraint for any set of external operators, labelled by the indices $i_1, \ldots , i_4$. One now uses the Virsoro fusion kernel of Ponsot and Teschner \cite{Ponsot:1999uf} to express the s-channel blocks in terms of their `t-channel' cousins,
\begin{equation}
{\cal F}_t (P_t | 1-z)= \int\limits_{\cal C} \frac{dP_s}{2} \mathbb{F}_{P_t P_s}\left[ \begin{array}{cc}
    3 & 4  \\
    2 & 1
\end{array}\right]  {\cal F}_s (P_s |z )\,,
\end{equation}
where the contour ${\cal C}$ to be chosen depends on the heavy states $i_1 \ldots i_4$, on which the blocks as well as the Virasoro fusion kernel depend. If the external operators are sufficiently heavy the contour may be chosen, such that the integral above becomes
\begin{equation}\label{eq.HeavyInversion}
{\cal F}_t (P_t | 1-z)= \int\limits_{\mathbb R}\frac{dP_s}{2} \mathbb{F}_{P_t P_s} \left[ \begin{array}{cc}
    3 & 4  \\
    2 & 1
\end{array}\right]{\cal F}_s (P_s |z )\,, \qquad (\textrm{heavy} \,\, \textrm{externals})
\end{equation}
for which $P_s \in \mathbb{R}$. For generic external operators below threshold, `sufficiently heavy' means that their combined weight puts us  in the continuum of heavy states $(h, \bar h) > \frac{c-1}{24} $. This is the case if they satisfy the following pair-wise conditions \cite{Ponsot:1999uf}, most simply stated in terms of the Liouville parameters $\alpha$ introduced above,
\begin{equation}\label{eq.PairwiseConditions}
    {\rm Re} (\alpha_1 + \alpha_2) > \frac{Q}{2}, \qquad  {\rm Re} (\alpha_3 + \alpha_4) > \frac{Q}{2}\,\qquad {\rm Re} (\alpha_1 + (Q-\alpha_2)) > \frac{Q}{2}\,, \ldots
\end{equation}
where the dots stand for five further permutations of the first two conditions, where we replace $\alpha_i \rightarrow Q-\alpha_i$. These conditions can be satisfied by placing the external operators in the continuum range of heavy states, or by considering pairwise sub-threshold operators, whose combined effect is nevertheless above threshold, i.e. by pairwise conditions,
\begin{equation}
    1 > \sqrt{1-\frac{24 h_1}{c-1}} + \sqrt{1-\frac{24 h_2}{c-1}}~,~ \sqrt{1-\frac{24 h_3}{c-1}} + \sqrt{1-\frac{24 h_4}{c-1}}\,\ldots \,.
\end{equation}
Of course all these are understood to come together with the analogous conditions for the anti-holomorphic weights.\footnote{For a theory of pure gravity our tensor model will naturally only involve such states in the sums, but more generally one may want to include additional light states below the continuum, in which case the choice of contour becomes more involved. This is explained, for example in \cite{Teschner:2001rv,Collier:2018exn}. } This condition comes from a careful consideration of the region of holomorphicity of the inversion kernel, \cite{Ponsot:1999uf,Collier:2018exn,Kusuki:2018wpa}, making sure that none of the towers of poles cross the contour in \eqref{eq.HeavyInversion}.
 If the combined weights of the external operators are in the discrete sub-threshold range, then additional discrete poles must be included in the contour integration. As we mentioned above, for the case of primary interest, namely pure gravity, the only sub-threshold states come from the identify module, and thus this subtlety does not arise.
As announced, the integral kernel $\mathbb{F}_{P_s P_t}$ is the Virasoro fusion kernel of Ponsot and Teschner. Together with its anti-holomorphic equivalent, it gives rise to a 6j symbol of Virasoro, which we define in equation \eqref{eq.6jDefition}.

There is an analogous transformation from one channel to another for the anti-holomorphic blocks, and we use it below. This allows us to implement the crossing of the conformal blocks in terms of the CFT data by equating the coefficient of each s-channel block\footnote{This assumes linear independence of the blocks themselves. In order to establish such statements, one should use the orthogonality of conformal partial waves, which are known under suitable conditions to form a mathematically complete basis. Statements involving conformal blocks are then obtained by contour deformation arguments. For results on rigorously establishing such inversion formulae in the case of the global conformal group, see \cite{Simmons-Duffin:2017nub}.} in \eqref{eq.BasicVirasoroCrossing}. This results in the expression
\begin{equation}\label{eq.crossconst}
\sum_{q}\left( C_{i_1 i_2 q}C_{i_3 i_4 q} \delta^{(2)} \left(P_s-P_q \right) -  C_{i_1 i_4 q}C_{i_2 i_3 q} \left| \mathbb{F}_{P_q P_s}\left[ \begin{array}{c c} P_3 & P_4 \\ P_2 & P_1 \end{array}\right]\right|^2  \right) = 0\,,
\end{equation}
where the absolute square of the crossing kernel, written here for clarity with all indices, gives rise to the Virasoro 6j symbol
\begin{equation}\label{eq.6jDefition}
    \begin{Bmatrix}  \mathcal O_q & \mathcal O_4 & \mathcal O_1 \\ \mathcal O_s & \mathcal O_2 & \mathcal O_3\end{Bmatrix} = \left\{ \begin{array}{c} \left|\rho_0(P_s)^{-1}C_0 (P_1, P_4, P_q) C_0 (P_2, P_3, P_q) \mathbb{F}_{P_q P_s} \left[ \begin{array}{c c} P_3 & P_4 \\ P_2 & P_1 \end{array}\right] \right|^2  \\[25pt]
   \left|\rho_0(P_q)^{-1}C_0 (P_1, P_2, P_s) C_0 (P_3, P_4, P_s) \mathbb{F}_{P_s P_q} \left[ \begin{array}{c c} P_3 & P_2 \\ P_4 & P_1 \end{array}\right] \right|^2
    \end{array}
    \right. \,.
\end{equation}
Note that we have defined the $6j-$symbol in both OPE channels and that the $6j-$symbol enjoys tetrahedal symmetry. The fusion kernels, written in the two channels enjoy the orthogonality relation
\begin{equation}
\label{eq.InversionOrtho}
    \int\limits_{\cal C} \left| \mathbb{F}_{P_p P_s} \left[ \begin{array}{c c} P_3 & P_4 \\ P_2 & P_1 \end{array}\right]\right|^2  \,\left| \mathbb{F}_{P_s P_q} \left[ \begin{array}{c c} P_3 & P_2 \\ P_4 & P_1 \end{array}\right]\right|^2 \frac{d^2 P_s}4 = \delta^{(2)} \left(P_p - P_q \right)\,,
\end{equation}
which we will make use of shortly.

As explained above, the origin of \eqref{eq.crossconst} is the expansion of the crossing equation in the basis of s-channel principal series blocks, and so should be understood in the sense of a distribution that can be integrated against a function of $P_s$ with the same analyticity properties as the conformal blocks. For Virasoro representations, the principal series includes the physical (real weight) operators above the $(c-1)/24$ threshold, so for such $P_q$ the delta function is the usual Dirac delta. For below threshold operators, including the identity, it can be defined using contour manipulations in the complexified $P_s$ plane. 

Note that \eqref{eq.crossconst} is the crossing equation projected down to a single conformal block. This can be obtained from the usual crossing equation as a function of cross-ratios, by integrating it against an appropriate function. To literally project down to a single conformal-block, this would require knowledge of the four-point function with arbitrary precision in Lorentzian kinematics, which we do not have in approximate CFTs due to the kinematic limitation $\zl$. Because of this, the delta function should be thought of as smeared into a smooth function of $P_q$ whose width around $P_s$ is much less than 1 but much greater than the microscopic level spacing. This is where $\zl$ enters in the definition of the tensor model, although we will not make the connection explicit here. One can then imagine a triple scaled limit of the tensor model, obtained by taking the size of the tensors and the coefficient of the potential to infinity while sending the width of the smearing function to zero. 

Now to derive the potential for the tensor model, we note that in the truncated space of our approximate CFT data, we view the four-point crossing equation \eqref{eq.crossconst} as a tensor constraint of the form
\begin{equation}
M_{i_3 i_4}^{i_1 i_2} (P_s, \bar P_s) = 0\,,
\end{equation}
which we therefore add to the tensor model potential as the term
\begin{equation}
V_{4} = \sum_{i_1\cdots i_4}{\!\!}'\int\limits_{\mathbb R \cup \mathds1} \left| M_{i_3 i_4}^{i_1 i_2}(P_s, \bar P_s) \,  \right|^2 |\mu(P_s) |^2 \frac{d^2P_s}4  \,,
\end{equation}
where, $\mathbb R \cup \mathds1$ in the limits of the integration represents that the identity is included for the corresponding operator integral and $\mu(P_s)$ is a measure chosen to make the above integral into an orthogonal inner product on the inversion kernel,
\begin{equation}
    \mu (P_s) = \frac{1}{\rho_0(P_s) C_0(P_1, P_2, P_s) C_0 (P_3, P_4, P_s)}\,.
\end{equation}
Using orthogonality of the fusion kernels we arrive at the quartic potential\footnote{To avoid cluttering lengthy expressions, we will use the shorthand notation $C_0(ijk)\equiv C_0(P_i,P_j,P_k)$ and $\rho_0(i)=\rho_0(P_i)$ from now on.}
\begin{eqnarray}\label{eq.V4}
V_4 &=& 2 \sum_{i_1\cdots i_4}{\!\!}' \sum_{p,q}\left( \frac{C_{i_1 i_2 p} C_{i_3 i_4 p} C_{i_1 i_2 q} C_{i_3 i_4 q}}{|\rho_0(p)C_0(12p)C_0(34p)|^2} \delta^{(2)} \left(P_p - P_q \right) - \right. \nonumber \\ &&  \hspace{3.5cm}\left. \frac{C_{i_1 i_2 p} C_{i_3 i_4 p} C_{i_1 i_4 q} C_{i_3 i_2 q}}{|C_0(12p) C_0 (34p) C_0(32q) C_0 (14q)|^2 } \begin{Bmatrix}
    \mathcal O_q & \mathcal O_{4} & \mathcal O_1 \\
    \mathcal O_p & \mathcal O_2 & \mathcal O_3
\end{Bmatrix} \right)\,.
\end{eqnarray}
Our procedure has thus produced a quartic contribution to the potential of the tensor model. Note that the sum over intermediate states $(p,q)$ also includes the identity (and potentially other sub-threshold states, for non-minimal versions of the tensor model). Moreover, we have not included factors of the density of states in the expression for the potential, since the sums already account for any potential degeneracies. The identity contribution will naturally produce a quadratic kinetic term for the tensors, as we shall see in section \ref{sec43}.
\begin{figure}[tb]
    \centering
    \includegraphics[width=.8\textwidth]{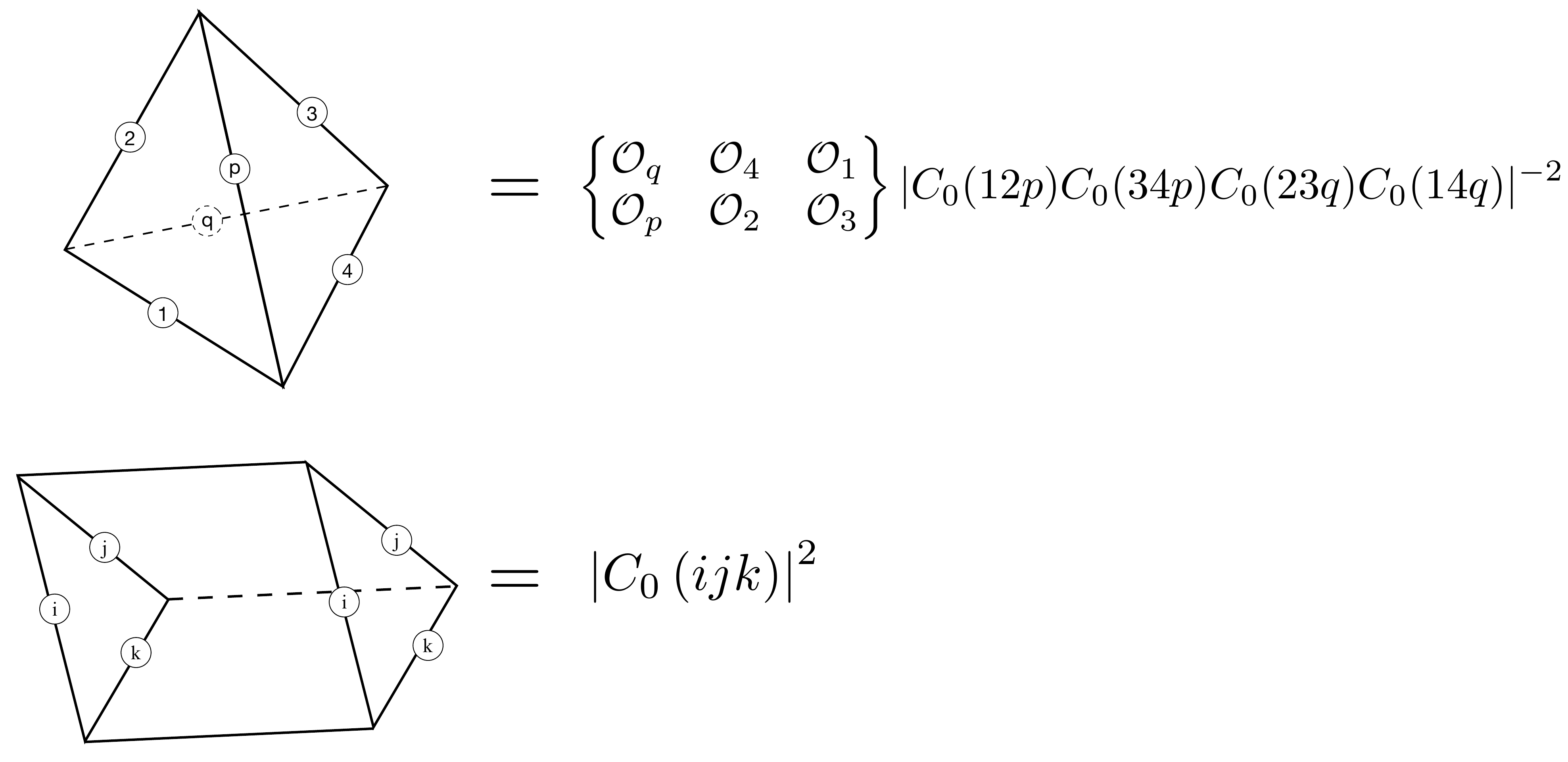}
    \caption{The Feynman rules of the tensor model after resummation. Each quartic vertex is given by a three-simplex, weighted by a Virasoro $6j-$symbol. The propagator glues two faces of these simplices and evaluates to the `$C_0$ formula'. As we show below this re-summed two-point function gives rise to the same quadratic moment as the Gaussian model of \cite{Chandra:2022bqq}. Interestingly the bare model, before resummation, has the same quadratic and quartic vertices, but with different uphysical $a$-dependent coefficients, and moreover possesses another four-valent vertex with a different index-structure, which we do not show here (see Equation \eqref{eq.V4}). A similar comment holds for the quadratic terms (see Equation \eqref{eq.TensorPropagator}).}
    \label{fig.TensorFeynman rules}
\end{figure}
The Virasoro $6j$ symbol possesses tetrahedral symmetry, and in fact it is natural to associate it with a 3-simplex whose edges are labelled by the external Virasoro highest-weight representations associated to the indices of the structure constants in the potential. Together with the propagator, which glues together two faces of such 3-simplices (we derive the exact expression of the propagator below), this gives rise to the graphical Feynman rules in Figure \ref{fig.TensorFeynman rules}. A typical diagram contributing to observables of the model thus constructs a discrete three-dimensional Euclidean manifold, in a fashion analogous to the `simplicial' approach to quantum gravity (see e.g. \cite{Hartle:2022ykc,Williams:1991cd}). 

Two major differences with previous constructions are that 1) we use representations of Virasoro to label edges and 2) we also integrate over the conformal dimensions themselves, in the form of the matrices of left- and right-moving conformal weights. As a consequence it is an intriguing possibility that our matrix-tensor model bridges ideas from simplicial gravity and the more recent work on VTQFT of \cite{Collier:2023fwi}. We shall now proceed to determine the other part of the potential coming from modular crossing of torus one-point functions.

\subsection{Modular crossing constraints and spectral density}
A two-dimensional CFT satisfies further crossing-type constraints beyond the four-point crossing on the sphere we discussed in the previous section. We will now discuss the implication of modular covariance of torus one-point functions, in particular concentrating on the S-modular transformation.

Following the arguments we made above for the crossing kernel of the conformal blocks on the sphere, we also have the modular covariance of the one point functions on a torus of modular parameter $\tau, \bar \tau$
\be
\braket{O}_{-\frac{1}{\tau}}= \tau^h\bar{\tau}^{\bar{h}} \braket{O}_{\tau} \,,
\ee
which implies the following relationship on the CFT data, 
\begin{equation}\label{eq.modular_crossing}
    \tilde \rho [P'|\mathcal O_0] = \int\limits_{\cal C} \frac{d^2P}4 \left| \mathbb S_{P'P}[P_0]\right|^2 \rho[P|\mathcal O_0]~,
\end{equation}
where, $\mathcal O_0$ is the operator inserted on the torus. The integration contour ${\cal C}$ can be chosen to be the same as above, under less restrictive conditions as \eqref{eq.PairwiseConditions}. In fact, as is shown in \cite{Teschner:2003at}, and explained for example in the appendix of \cite{Collier:2019weq}, none of the poles appearing in the S-modular kernel can cross the integration contour ${\cal C}$, so long as all external weights are in the unitary range.  The $\rho$'s are the coarse-grained weighted densities of states defined in the dense part of the spectrum,
\begin{align}
    \rho[P|\mathcal O_0] &= \sum_{\mathcal{O}} C_{\mathcal O \mathcal O \mathcal O_0} \delta^{(2)}(P-P_{\mathcal O}) \,, \label{eq.rho.def}\\
    \tilde\rho[P'|\mathcal O_0] &= \sum_{\mathcal{O'}} \tilde C_{\mathcal O' \mathcal O' \mathcal O_0} \delta^{(2)}(P'-P_{\mathcal O'})~, \label{eq.rho-tilde.def}
\end{align}
where, $P,P'$ are the Liouville momenta corresponding to the operators $\mathcal O, \mathcal O'$, respectively.
In the presence of light operators, one can include a sum over the corresponding delta functions in the density of states.
Here, we are using the $\tilde \cdot$ to denote the quantities in the crossed modular channel corresponding to the modular parameters $-1/\tau,-1/\bar\tau$, as opposed to the un-tilded  quantities in the channel corresponding to modular parameter $\tau, \bar \tau$. We will drop the tilde subsequently since this difference is only notational and the OPE coefficients as well as the density of states are not channel dependent. Substituting the expressions for $\rho, \tilde \rho$ from \eqref{eq.rho.def}, \eqref{eq.rho-tilde.def} into \eqref{eq.modular_crossing} one gets,
\begin{equation}\label{eq.ModularConstraint}
\sum_i C_{ii j} \delta^{(2)} \left(P - P_i \right)  - C_{iij} \left| \mathbb{S}_{P P_i}\left[ P_j\right]\right|^2 = 0\,,
\end{equation}
We view this again as an equation $M_j(P_q, \bar P_q) :=0$, which we enforce via its absolute square in the potential
\begin{equation} \label{VS}
V_{\rm S} = \sum_{j} \int\limits_{\cal C}\! \frac{d^2P}4 |\mu_S(P)|^2 \left| M_{j}(P, \bar P) \right|^2 \,.
\end{equation}
The sum/integral over the Liouville momentum includes all the operator insertions on the torus. The measure $\mu_S$ is chosen so as to make the above into an orthogonal inner product and is given by,\footnote{This measure is only applicable to the continuous part of the Liouville spectrum, i.e. for $h,\bar h\ge \frac {Q^2}4$, but not when the external insertion is $\mathcal O_j = \mathds 1$.}
\begin{equation}
\label{eq.measureS}
    \mu_S(P_q) = \rho_0(q)C_0(qqj)~.
\end{equation}
Moreover, as mentioned when discussing four-point crossing, equation \eqref{VS} should be understood in a smeared sense.

Before we proceed, let us list some identities of the modular inversion kernel that we need in subsequent computations,

\begin{align}
\label{eq.SXchangeSymm}
    \mathbb S_{PP'}[P_0] \rho_0(P) C_0(P,P,P_0) &= \mathbb S_{P'P}[P_0] \rho_0(P') C_0(P',P',P_0) \\
\label{eq.SOrtho}
    \int\limits_{\mathbb R}\!\!\frac{dP}{2}\mathbb S_{PP_1}[P_0] \mathbb S_{P_2P}[P_0] &= e^{i \pi \Delta_0} \delta(P_1-P_2)~.
\end{align}
When the inserted operator $\mathcal O_0 = \mathds 1$,
\begin{align}
\label{eq.SXchangeSymmId}
	\lim_{P_0\to\mathds1} C_0(P_1,P_2,P_0) &\to \rho_0^{-1}(P_1)\delta(P_1-P_2)\\
	\lim_{P_0\to\mathds1} \mathbb S_{PP'}[P_0] \rho_0(P) C_0(P,P,P_0)  = \mathbb S_{PP'}[\mathds1] \delta(P-P') &= \mathbb S_{P'P}[\mathds1] \delta(P-P') \nonumber \\
														    & = \lim_{P_0\to\mathds1}\mathbb S_{P'P}[P_0] \rho_0(P') C_0(P',P',P_0) \\
\label{eq.SOrthoId}
    \int\limits_{\mathbb R}\!\!\frac{dP}{2}\mathbb S_{PP_1}[\mathds1] \mathbb S_{P_2P}[\mathds 1] &= \delta(P_1-P_2)~.
\end{align}
Additionally, the inversion kernels we need subsequently are given by \cite{Collier:2019weq},
\begin{align}
\label{eq.SIdins}
    &\mathbb S_{PP'}[\mathds 1] = 2\sqrt2 \cos(4\pi PP')\\[5pt]
\label{eq.SIdinv}
     \rho_0(P) := \mathbb S_{P\mathds 1}[\mathds 1] = \mathbb S_{P,\frac i2 \left(b+\frac1b\right)}[\mathds 1] &- \mathbb S_{P,-\frac i2\left(b-\frac1b\right)}[\mathds 1] = 4\sqrt 2 \sinh(2\pi bP) \sinh\left(2\pi \frac Pb\right)~.
\end{align}

The terms that correspond to the $\mathcal O_j = \mathds 1$ operator impose modular invariance of the torus without any operator insertion and need to be treated separately,
\begin{align}
\label{eq.VSIdentOld}
	V_{\rm S,\mathds 1} &= \int\!\frac{d^2P}{4} \sum_{i,k} \left( \delta^{(2)}(P-P_i) - \left| \mathbb S_{PP_i}[\mathds 1]\right|^2\right)\left( \delta^{(2)}(P-P_k) - \left| \mathbb S_{PP_k}[\mathds 1]\right|^2\right)~.
\end{align}
In writing the above expression we have used the normalisation, $C_{ii\mathds 1} = 1$.

Let us briefly pause to point out that the identity contribution of the potential enforces the Cardy density of states as a saddle-point equation. To see this we only need to notice that the saddle-point conditions coming from $V_{S,\mathds 1}$ are solved precisely by satisfying \eqref{eq.ModularConstraint}, which is of course happening by design. As has been shown, for example in \cite{Collier:2019weq}, this condition can be rewritten in the form
\begin{equation}
    |\rho(P)|^2 = \left| \mathbb S_{P \mathds 1}[\mathds 1] \right|^2 + \sum_i \left| \mathbb S_{P P_i}[\mathds 1] \right|^2 \sim 2 e^{2\pi Q (P + \bar P)}\,,
\end{equation}
where in the last step we have exploited the fact that the identity contribution dominates.\footnote{This is true in any CFT asymptotically for $P\rightarrow\infty$, but if a sparse spectrum below dimension $c/12$ is assumed (which is certainly the case here since we have no operators at all below the threshold), the formula holds at large-$c$ for all operators above threshold \cite{Hartman:2014oaa}.} It is elementary to use the definitions given at the start of section \ref{sec.tensor} to convert from Liouville notation to more standard conformal weights, $(h, \bar h)$, in order to convince oneself that this is indeed the usual Cardy density of states $|\rho_0(h)|^2  \sim e^{2\pi\sqrt{\frac{c}{6}\left(h - \frac{c}{24}\right)}} e^{2\pi\sqrt{\frac{c}{6}\left(\bar h - \frac{c}{24}\right)}}$.

Getting back to the modular potential, further careful treatment of the expression $V_{S, \mathds 1}$ leads to the following potential, 
\begin{align}
\label{eq.VSIdentExp}
	\frac{ V_{\rm S, \mathds 1} }2 & = \sum_{i,k}{\!}^{'}\delta^{(2)}(P_k-P_i) - 8 \sum_{i,k}{\!}^{'}|\cos(4\pi P_iP_k)|^2 
			     -64 \sum_i{\!}^{'}\left|\sinh(2\pi b P_i)\sinh\left(2\pi\frac {P_i}b\right)\right|^2 + \ldots
\end{align}
Here, $\sum{\!}'$ denotes the sum over the operators/states excluding the identity operator.
Furthermore, ``$\cdots$'' in \eqref{eq.VSIdentExp} represents terms that do not depend on the physical operators $\mathcal O_i$ and are therefore just a normalisation factor for our purposes.
In arriving at the above result we have used the following relationships:
\begin{align}
	\int \!\frac{d^2P}{4} \, \delta^{(2)}(P-\mathds 1) |\mathbb S_{PP_i}[\mathds 1]|^2 &= |\mathbb S_{\mathds 1P_i}[\mathds1]|^2 \\
	\int \!\frac{d^2P}{4} \, |\mathbb S_{PP_i}[\mathds 1]|^2 |\mathbb S_{P\mathds 1}[\mathds 1]|^2 &= \delta^{(2)}\left(P_i \pm \frac i2\left(b + \frac1b\right)\right) - \delta^{(2)}\left(P_i \pm \frac i2\left(b - \frac1b\right)\right)~\\
	&=: \delta^{(2)}(P_i-\mathds1) .\nonumber
\end{align}
Note that the delta function $\delta^{(2)}(P-\mathds1)$ is defined by the corresponding sum over four individual delta functions because of its degenerate nature. In the first equation, we have interpreted the $P$ integral as an off-contour delta function, and hence this term contributes to the last term in \eqref{eq.VSIdentExp}. The delta-function on the RHS of the second equation above is a sum over two different delta-functions for each possible choice of signs. Since we are considering an explicit sum over above-threshold states corresponding to Liouville momentum $P_i$, the RHS is 0 for the states that we are considering. However, the contribution of this term could be more subtle when the delta functions in the above equations are considered as a smeared version of the distributional function with a width around $P_i$ which is much less than 1 but much greater than the microscopic level spacing. We leave this more careful analysis with smeared crossing kernels for future work.

The remaining terms corresponding to the external operator $\mathcal O_j \neq \mathds 1$ can be worked out explicitly to obtain,
\begin{align}
\label{eq.VSIdentOLD}
V_{\rm S, \cancel{\mathds 1}} &= 2\sum_{i,j,k}{\!}'\left| \rho_0(i)C_0(iij)\right|^2 C_{iij}C_{kkj} \left( \delta^{(2)} \left(P_i - P_k \right)   - \left| \mathbb{S} \left[{\cal O}_j \right]_{P_i P_k} \right|^2  \right)\,. 
\end{align}
In arriving at this expression, we have used the orthogonality relation of the $\mathbb{S}$-modular kernel. The total modular potential is the sum of the two terms computed above,
\begin{equation}
    V_{\rm S} = V_{\rm S,\mathds 1} + V_{\rm S,\cancel{\mathds 1}}~.
\end{equation}
So far, we have not imposed integrality of spin in the definition of the matrices $L_0$ and ${\bar L}_0$, thus T modular invariance is a non-trivial constraint. Following the general logic, we should add its square to the matrix potential. Note that only torus 0-point function T invariance is required, as this implies that all spins are integers and there are no further constraints. The condition is that 
$$\frac{e^{2\pi i (\tau+\bar{\tau})\frac{1-c}{24}}}{\eta(\tau)\eta(\bar\tau)}\sum_i e^{2\pi i\tau h_i}{e}^{{2\pi i\bar{\tau}\bar h}_i} (1 - e^{2\pi i(h_i - \bar{h}_i)}) =0 \,.$$

The local version of the constraint, obtained from an inverse Laplace transform, is $\sum_i \delta(h_i - h) \delta(\bar{h}_i - \bar{h}) \sin(\pi(h_i -\bar{h}_i)) = 0$ \,. 

Interpreting the delta function as smeared, the resulting potential given by squaring the equation would be double trace. However, as integral spins are an exact feature of microscopic CFTs, it might be more natural to treat the delta function as applying microstate by microstate, which results in the addition of the single trace potential $$V_T = \Tr\sin^2(\pi(L_0 - \bar{L}_0)).$$ It is obvious that this vanishes exactly when the spins are integral. 

\subsection{The total potential, simplicial propagator \label{sec43}}
Imposing modular crossing of torus one-point functions and four-point crossing on the sphere then gives us the following potential of our tensor/matrix model
\begin{equation}
\tfrac{1}{2}V = V_{\rm S} + g_4 V_{4}\,. 
\end{equation}
The potential contains terms that only restrict the matrix part of the potential, terms quadratic in the tensors, as well as the quartic interaction among tensors controlled by the Virasoro 6j symbols. In order to understand the propagator of the tensors, let us collect all terms quadratic in the tensors into $V_2 \subset \frac12 V$, giving
\begin{align}
\label{eq.TensorPropagator}
    V_2 &= 2\sum_{i,j,k}{\!}'\left| \rho_0(i)C_0(iij)\right|^2 C_{iij}C_{kkj} \left( \delta^{(2)} \left(P_i - P_k \right)   - \left| \mathbb{S} \left[{\cal O}_j \right]_{P_i P_k} \right|^2  \right) \nonumber \\
    &\hspace{3cm}- 4g_4\sum_{i,j,k}{\!}' \frac{C_{ijk}C_{ijk}}{|C_0(ijk)|^4} \begin{Bmatrix}
        \mathcal O_k & \mathcal O_j & \mathcal O_i \\
        \mathds 1 & \mathcal O_i & \mathcal O_j
    \end{Bmatrix} \nonumber \\
    &= 2\sum_{i,j,k}{\!}'\left| \rho_0(i)C_0(iij)\right|^2 C_{iij}C_{kkj} \left( \delta^{(2)} \left(P_i - P_k \right)   - \left| \mathbb{S} \left[{\cal O}_j \right]_{P_i P_k} \right|^2  \right) \nonumber \\
    &\hspace{3cm}- 4g_4\sum_{i,j,k}{\!}' \frac{C_{ijk}C_{ijk}}{|C_0(ijk)|^2}  \,.
\end{align}
The first term comes from the S-modular potential, which is naturally quadratic in the tensors, and explicitly excludes the identity contribution on any of the summed indices. The second term comes from $V_4$, for the special case that indices are pairwise the same in one channel. In this case (e.g. $i_1 = i_4 := i$, $i_2 = i_3 = j$) we isolate the contribution from both pairs fusing into the identity (noting that then $C_{ii \mathds{1}} = 1$). We have also used the identity,
\begin{equation}
	\begin{Bmatrix}
		\mathcal O_k & \mathcal O_j & \mathcal O_i \\
    		\mathds 1 & \mathcal O_i & \mathcal O_j
	\end{Bmatrix} = |C_0(ijk)|^2 \,.
\end{equation}
to simplify the second term.

\subsection{Going from potential to moments in the tensor model}\label{sec.fullbare}
We have now defined our tensor model, which contains quadratic and quartic order terms in its potential. From the quadratic terms we were able to read off that the bare propagator corresponds  to the $C_0$ formula found previously, which seems reassuring. It is important to note, however, that this agreement is somewhat arbitrary in the sense that what really can be compared to section \ref{sec.var-cross} (and the ensemble of \cite{Chandra:2022bqq}) is the full two-point function, computed from our matrix/tensor potential. 

Similarly, the connected four-point function obtained by gluing four bare propagators to the bare tetrahedral vertex evaluates, up to an overall factor, to the 6j symbol itself, which is precisely the quartic moment \eqref{eq.4cnonGauss} that we found was required by the variance of the crossing equation. It is only after resummation of infinitely many diagrams that the overall inverse factors of the coefficient $a$ in front of the potential disappear. 

Note that what we are summing here is a certain class of diagrams, analogous to the leading planar order in an 't Hooft expansion. The above four point amplitude is precisely the contribution of the four-boundary wormhole in 3d gravity as evaluated by the methods of VTQFT \cite{Collier:2023fwi}. The complete amplitude in the tensor model includes further contributions that we expect are associated to more complicated topologies in 3d gravity.

We will perform this computation below by solving the appropriate Schwinger-Dyson equations of the model. Before doing so, it is worth illustrating a subtlety in the definition of the model, stemming from the fact that the model was defined in the first place by `squaring a constraint'. 
\subsection*{Propagator vs. variance in `constraint-squared' potentials}
 Before going to the full tensor model, let us illustrate this behavior in a (much) simpler example. Full details on this computation may be found in appendix \ref{app.ToyModel} for the interested reader. 

Suppose, for the purpose of illustration, that we want to define a model whose saddle-point equation corresponds to solving the constraint $f(|\phi|)=0$ for a vector valued variable $\phi_i$. For reasons that should be clear from the above construction of the matrix-tensor model, we are interested in quadratic constraints, $\phi^2 = C_0^2$ for some constant $C_0$. We can implement this by squaring the constraint, to obtain
\begin{equation}
  Z_1 =   \int [d\phi] e^{-N (\phi^2 - C_0^2)^2}\,,
\end{equation}
but it should be clear that this construction is not unique. An equally good choice would have been the following model
\begin{equation}
    Z_2 =   \int [d\phi] e^{-N \left(\frac{\phi^2}{C_0^2} - 1\right)^2}\,,
\end{equation}
which evidently gives rise to the same quadratic constraint at large-$N$.
While the difference may seem innocuous, the two actually lead to drastically different perturbation theories, which we can see already by noting that the bare `propagators' are
\begin{equation}
    G^{\rm bare}_1 = - 1/4C_0^2 \qquad  G^{\rm bare}_2 = -C_0^2/4\,,
\end{equation}
where in both cases we have adopted the normalization $N^{-1}\phi^T \phi$ for the two-point function, bare or resummed. One notices therefore that in both cases the bare propagator gives the wrong answer -- the right answer, by definition would be the straight solution of the quadratic constraint. In the first case we get something proportional to the inverse of the correct answer, while in the second case, the sign and the pre-factor are incorrect. It should be clear that we could invent other forms of the potential which result in yet more `incorrect' answers, as compared to the straightforward solution of the quadratic constraint.

Of course, the bare propagator is not what we should compare to the second moment of the distribution, instead the latter corresponds to the full two-point function in the interacting theory. In this simple toy model it is clear that in both cases this is given by evaluating the $ \left\langle N^{-1}\phi^T \phi \right\rangle$ on the quadratic constraint, which leads to the same answer in both models.
\begin{equation}
    G^{\rm full}_1 = C_0^2 \qquad  G^{\rm full}_2 = C_0^2\,,
\end{equation}
It is instructive to understand in detail how this result comes about in a resummed perturbative evaluation of the two-point function in each model, which we perform in appendix \ref{app.ToyModel}. The important take-away lesson for the tensor model is that in order to compare with definitions of the OPE statistics (and matrix-element) statistics, as defined in terms of moments of the distribution -- e.g. in our section \ref{sec.var-cross} or in \cite{Chandra:2022bqq}, care needs to be taken to go from the probability distribution to the moments, e.g. by solving the hierarchy of Schwinger-Dyson equations. We note that this issue already came up and was addressed in \cite{Jafferis:2022uhu,Jafferis:2022wez} for two-dimensional gravity. As a final comment before moving on to the Schwinger-Dyson analysis of the matrix/tensor model, let us point out that with the most natural definition of the constraint-squared potential (which we have chosen), the only difference between the bare and full two-point function is a renormalized coefficient, so that at first glance the issue may not seem too dramatic. We emphasize, however, that going to the full Schwinger-Dyson analysis is necessary to show the in-detail compatibility of our results in section \ref{sec.var-cross} and the present section, and that otherwise, the non-uniqueness of the potential and resulting perturbation theory described above might seem puzzling.

\subsection*{Exact two- and four-point function: Schwinger-Dyson approach}
The exact resummed two-point function of the tensor model satisfis a Schwinger-Dyson equation, which follows from the identity
\begin{equation}\label{eq.TensorSDIdentity}
    \int\!\!D[C]~ \frac\partial{\partial C_{ijk}} \Big( C_{lmn}\; e^{-aV[C]} \Big) = 0.
\end{equation}
Note that for clarity, we will suppress the dependence of the above integral on the variables $h,\bar h$ which are not relevant for the current discussion.
Also, we suppress the terms in the potential $V[C]$ that do not depend on the OPE coefficients,
\begin{align}
    \frac{V[C]}2 &= \frac{V_{S,\mathds 1}}2 + \sum_{i,j,k}{\!}'\left| \rho_0(i)C_0(iij)\right|^2 C_{iij}C_{kkj} \left( \delta^{(2)}_{i,k} - \left| \mathbb{S}_{P_i P_k} \left[{\cal O}_j \right] \right|^2  \right) \nonumber \\
    &\hspace{0.125cm}- 2g_4\sum_{i,j,k}{\!}' \frac{C_{ijk}C_{ijk}}{|C_0(ijk)|^4} \begin{Bmatrix}
        \mathcal O_k & \mathcal O_j & \mathcal O_i \\
        \mathds 1 & \mathcal O_i & \mathcal O_j
    \end{Bmatrix} 
    + g_4 \sum_{\substack{i_1\cdots i_4\\p,q}}{\!\!}'\left( \frac{C_{i_1 i_2 p} C_{i_3 i_4 p} C_{i_1 i_2 q} C_{i_3 i_4 q}}{|\rho_0(p)C_0(12p)C_0(34p)|^2} ~ \delta^{(2)}_{p,q}   \right. \nonumber \\ 
    &  \hspace{4.5cm}\left.- \frac{C_{i_1 i_2 p} C_{i_3 i_4 p} C_{i_1 i_4 q} C_{i_2 i_3 q}}{|C_0(12p) C_0 (34p) C_0(23q) C_0 (14q)|^2 } \begin{Bmatrix}
        \mathcal O_q & \mathcal O_{4} & \mathcal O_1 \\
        \mathcal O_p & \mathcal O_2 & \mathcal O_3
    \end{Bmatrix} \right)~.
\end{align}
The term $V_{S,\mathds1}$ does not depend on the OPE coefficient and is not relevant for our purpose at this moment. Substituting this action in the identity \eqref{eq.TensorSDIdentity}, we generate the Schwinger-Dyson equation
\begin{align}\label{eq.SDexact}
    \frac{\delta^{ijk}_{lmn}}{2a} &= 2 \delta_{ij}\sum_b{\!}'\left| \rho_0(i)C_0(iik)\right|^2 \left\langle C_{lmn} C_{bbk} \right\rangle \left(\delta^{(2)}_{i,b}-\left| \mathbb{S}_{P_i P_b} \left[{\cal O}_k \right] \right|^2 \right)+ ~{\rm S_3 ~ perm.} \nonumber \\
    &\qquad -24 g_4 \frac{\left\langle C_{ijk} C_{lmn} \right\rangle}{|C_0(ijk)|^2} + 4g_4 \left(\sum_{i_1i_2}{\!}' \frac{\left\langle C_{lmn} C_{i_1i_2k} C_{ijk} C_{i_1i_2k} \right\rangle}{|\rho_0(k) C_0(ijk)C_0(12k)|^2} + {\rm ~ S_3 ~ perm.} \right) \nonumber \\
    &\qquad - 4g_4 \Bigg( \sum_{i_1,i_2,i_3}{\!\!\!}' \frac{\left\langle C_{lmn} C_{i_1i_2k} C_{ii_2i_3} C_{ji_1i_3} \right\rangle}{|C_0(ijk)C_0(12k)C_0(i23)C_0(j13)|^2} \begin{Bmatrix}
        \mathcal O_{3} &\mathcal O_{2} &\mathcal O_{i}\\
        \mathcal O_{k} &\mathcal O_{j} & \mathcal O_{1}
    \end{Bmatrix} + ~{\rm S_3 ~ perm.} \Bigg)~.
\end{align}
\begin{figure}[h!]
    \centering
  \includegraphics[width=\textwidth]{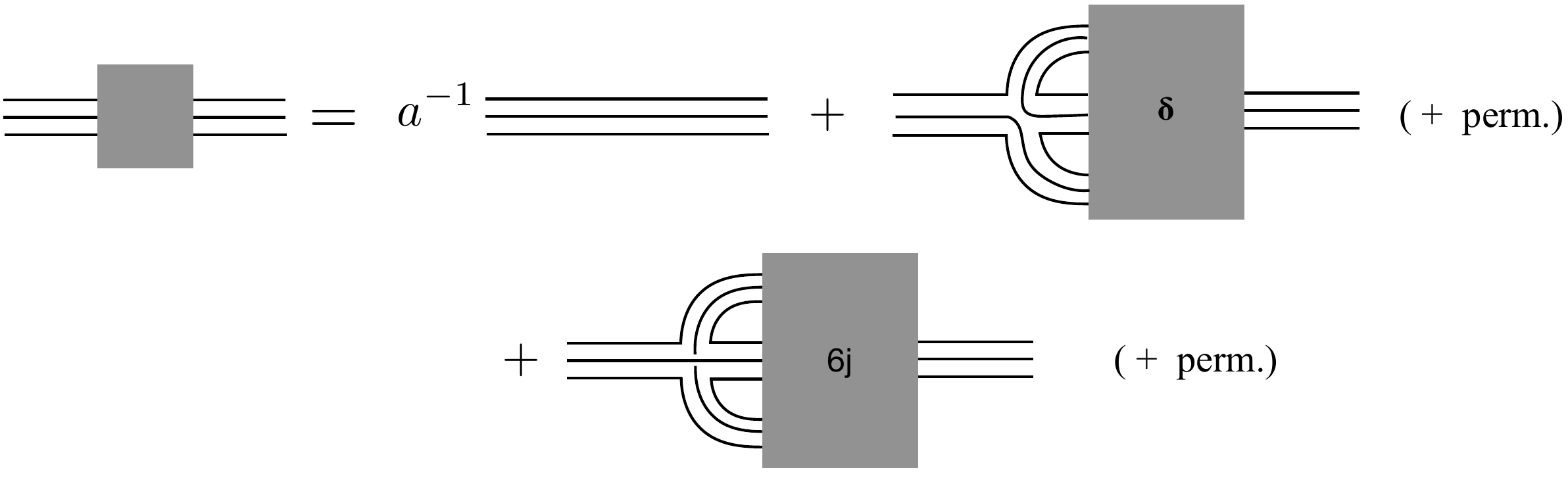}
    \caption{Diagrammatic representation of the Schwinger-Dyson equation \eqref{eq.SDsimple}. The left-hand side is the exact two-point function $\langle C_{ijk}C_{ijk} \rangle$, and the first term on the right-hand side is given by the bare propagator, which is suppressed at large $a$. The remaining terms are the two contributions due to the exact four-point function in different index contractions appearing in \eqref{eq.SDsimple}, with the second quartic vertex given by the $6j$ symbol. Note that we have chosen a `triple-line' notation in analogy with 't Hooft's double line notation for matrix models, but have suppressed the index labels on the lines to avoid clutter. Each index on the tensor $C_{ijk}$ corresponds to one of the three lines. Note that if the triple lines are condensed into one, the two different quartic vertices become indistinguishable, and consequently the SD equation here is structurally identical to the one of the toy model of appendix \ref{app.ToyModel}.}
    \label{fig.TensorSDEquation}
\end{figure}
In the above expressions, the permutations corresponds to the permutations of the $i,j,k$ indices. We consider the case $\{l,m,n\} = \{i,j,k\}, ~ i\neq j\neq k$ for the purpose of demonstrating the general pattern. To leading order at large $a$, the LHS of the Schwinger-Dyson equation can be taken to zero. Additionally, in this limit the spectrum of the theory can be approximated by a smooth density of states, as is standard in the study of RMT \cite{Meh2004}. The simplified expression is as follows,
\begin{align}
\label{eq.SDsimple}
0 &= - 24g_4 \frac{\langle C_{ijk} C_{ijk}\rangle}{|C_0(ijk)|^2} |\rho(i) \rho(j) \rho(k)|^2 \nonumber \\
      & \hspace{1cm} +4g_4 \int\limits_{i_1,i_2}{\!\!}^{'} \left| \rho(i)\rho(j)\rho(k)\rho(i_1)\rho(i_2)\rho(k)\right|^2\, \frac{\left \langle C_{ijk} C_{i_1i_2k} C_{ijk} C_{i_1i_2k} \right\rangle}{ \left|\rho_0(k) C_0(ijk) C_0(12k) \right|^2} + S_3~ {\rm perm.} \nonumber \\
    & \hspace{-0.65cm}-24 g_4 \!\!\!\! \int\limits_{i_1,i_2,i_3}{\!\!\!\!}^{'} \left| \rho(i)\rho(j)\rho(k)\rho(i_1)\rho(i_2)\rho(i_3)\right|^2\, \frac{\left \langle C_{ijk} C_{i_1i_2k} C_{ii_2i_3} C_{ji_1i_3} \right\rangle}{ \left| C_0(i23) C_0(j13) C_0(ijk) C_0(12k) \right|^2} \, \begin{Bmatrix}
        \mathcal O_3 & \mathcal O_2 & \mathcal O_i\\
        \mathcal O_k & \mathcal O_j & \mathcal O_1
    \end{Bmatrix}\,.
\end{align}
Since the last term on the RHS is inherently symmetric under the $S_3$ permutations of the indices $i,j,k$, we have already accounted for the contributions of these permutations in the overall factor of the term.
A graphical representation of the SD equation is given in Figure \ref{fig.TensorSDEquation}; one may note the structural similarity to our quartic toy model above.
One can now verify that the exact correlations that were used in the previous section, \eqref{Gaussianmoment}  and \eqref{eq.4cnonGauss}, satisfy these Schwinger-Dyson equations at leading order in the $a \gg 1$ limit, which is precisely the regime we are interested in.

The first term on the RHS of the Schwinger-Dyson equation \eqref{eq.SDsimple} evaluates to,
\begin{equation}
\label{eq.1stterm}
	\frac{\langle C_{ijk} C_{ijk}\rangle}{|C_0(ijk)|^2} |\rho(i) \rho(j) \rho(k)|^2 = \left|\rho_0(i) \rho_0(j) \rho_0(k)\right|^2~.
\end{equation}
In the large $a$ limit, we have replaced $\rho(i)\to\rho_0(i)$ which is the solution for the one-point function of the density of states. The correlation function appearing in the second term on the RHS of the above equation is given by (as discussed in the previous section),
\begin{align}
\label{eq.SDGauss}
\hspace{-0.28cm}\left\langle C_{ijk} C_{i_1i_2k} C_{ijk} C_{i_1i_2k} \right\rangle &= \left\langle C_{ijk} C_{ijk} \right\rangle \left\langle C_{i_1i_2k} C_{i_1i_2k} \right\rangle + \left\langle C_{ijk} C_{ijk} \right\rangle^2 \frac{\delta^{(2)}(P_i-P_{i_1})}{|\rho_0(i_1)|^2} \frac{\delta^{(2)}(P_j-P_{i_2})}{|\rho_0(i_2)|^2} \nonumber\\
& \hspace{4.2cm}+ \begin{Bmatrix}
        \mathcal O_k & \mathcal O_j & \mathcal O_i\\
        \mathcal O_k & \mathcal O_j & \mathcal O_1
    \end{Bmatrix} \frac{\delta^{(2)}(P_j - P_{i_2})}{|\rho_0(i_2)|^2}\\
\label{eq.SDGauss2}
								       &= |C_0(ijk)|^2 |C_0(12k)|^2 + |C_0(ijk)|^4 ~ \frac{\delta^{(2)}(P_i-P_{i_1})}{|\rho_0(i_1)|^2} \frac{\delta^{(2)}(P_j-P_{i_2})}{|\rho_0(i_2)|^2}\nonumber\\
& \hspace{4.2cm}+ \begin{Bmatrix}
        \mathcal O_k & \mathcal O_j & \mathcal O_i\\
        \mathcal O_k & \mathcal O_j & \mathcal O_1
    \end{Bmatrix} \frac{\delta^{(2)}(P_j - P_{i_2})}{|\rho_0(i_2)|^2} \,.
\end{align} 
The corresponding term in the above expression thereby simplifies to,\footnote{Note that we have already accounted for the permutations between the $i,j,k$ indices and therefore do not need to include them as different contribution in these contractions.}
\begin{align}
\label{eq.2ndterm}
	\int\limits_{i_1,i_2}{\!\!}^{'}& \left| \rho(i)\rho(j)\rho(k)\rho(i_1)\rho(i_2)\rho(k)\right|^2\, \frac{\left \langle C_{ijk} C_{i_1i_2k} C_{ijk} C_{i_1i_2k} \right\rangle}{ \left|\rho_0(k) C_0(ijk) C_0(12k) \right|^2} \nonumber \\
	&= \left| \rho_0(i) \rho_0(j)\rho_0(k) \right|^2 \int\limits_{i_1,i_2}{\!\!}^{'} \left| \rho_0(i_1)\rho_0(i_2) \right|^2 + \left|\rho_0(i) \rho_0(j) \rho_0(k)\right|^2 \nonumber \\
    & \hspace{2cm}+ \left| \rho_0(i)\rho_0(j)\rho_0(k)\right|^2 \int\limits_{i_1}{\!}^{'} \left|\rho_0(i_1)\right|^2\, \begin{Bmatrix}
        \mathcal O_k & \mathcal O_j & \mathcal O_i\\
        \mathcal O_k & \mathcal O_j & \mathcal O_1
    \end{Bmatrix} ~.
\end{align}
Note that the contribution of the second term in \eqref{eq.2ndterm} cancels that of the first term on the RHS of \eqref{eq.1stterm} along with all the $S_3$ permutations. The contribution of the first and the third term cancels the contribution of the last term on the RHS of \eqref{eq.SDsimple}. To see this let us simplify this term further using the expression for the correlation function,
\begin{align}
\label{eq.6jexactCorr.}
    \left\langle C_{ijk} C_{i_1i_2k} C_{ii_2i_3} C_{ji_1i_3} \right\rangle =& \begin{Bmatrix}
        \mathcal O_i &\mathcal O_j &\mathcal O_k\\
        \mathcal O_{1} &\mathcal O_{2} & \mathcal O_{3}
    \end{Bmatrix} \nonumber \\
    &\hspace{-1cm}+ \left( |C_0(ijk)C_0(1jk)|^4 ~ \frac{\delta^{(2)}(P_k-P_{i_3})}{|\rho_0(i_1)|^2} \frac{\delta^{(2)}(P_j-P_{i_2})}{|\rho_0(i_2)|^2} + S_3~{\rm perm.} \right)~.
\end{align}
The last term of \eqref{eq.SDsimple} using the above result for the 4-point function is,
\begin{align}
	\int\limits_{i_1,i_2,i_3}{\!\!\!\!\!}^{'}& |\rho(i)\rho(j)\rho(k)\rho(i_1)\rho(i_2)\rho(i_3)|^2 \frac{\left\langle C_{ijk} C_{i_1i_2k} C_{ii_2i_3} C_{ji_1i_3} \right\rangle}{|C_0(ijk)C_0(13j)C_0(i23)C_0(k12)|^2} \begin{Bmatrix}
        \mathcal O_i &\mathcal O_{2} &\mathcal O_{3}\\
        \mathcal O_{1} &\mathcal O_j & \mathcal O_k
    \end{Bmatrix} & \nonumber \\
						 &= \int\limits_{i_1,i_2,i_3}{\!\!\!\!\!}^{'}\left|{\rho_0(j) \rho_0(k) \rho_0(i_2) \rho_0(i_3)}\right|^2 \left| \mathbb F_{P_{i_1}P_i} \begin{bmatrix}
			P_j&P_k\\
			P_3&P_2
		\end{bmatrix}\mathbb F_{P_{i}P_{i_1}}
		\begin{bmatrix}
 			P_j&P_3\\
			P_k&P_2
		\end{bmatrix}\right|^2 \nonumber \\
       & \hspace{3cm}+ \left( \left| \rho_0(i)\rho_0(j)\rho_0(k)\right|^2 \int\limits_{i_1}{\!}^{'} \left|\rho_0(i_1)\right|^2\, \begin{Bmatrix}
        \mathcal O_k & \mathcal O_j & \mathcal O_i\\
        \mathcal O_k & \mathcal O_j & \mathcal O_1
    \end{Bmatrix} + S_3~{\rm perm.} \right) \nonumber \\
						 &= \left| \rho_0(j) \rho_0(k)\right|^2 \delta^{(2)}(P_i-P_i) \int\limits_{i_2,i_3}{\!\!\!}^{'} \left|\rho_0(i_2)\rho_0(i_3)\right|^2 \nonumber \\
       & \hspace{3cm}+ \left( \left| \rho_0(i)\rho_0(j)\rho_0(k)\right|^2 \int\limits_{i_1}{\!}^{'} \left|\rho_0(i_1)\right|^2\, \begin{Bmatrix}
        \mathcal O_k & \mathcal O_j & \mathcal O_i\\
        \mathcal O_k & \mathcal O_j & \mathcal O_1
    \end{Bmatrix} + S_3~{\rm perm.} \right)~. 
\end{align}
We have used the identities \eqref{eq.6jDefition} and \eqref{eq.InversionOrtho} to arrive at this result.
Note that the appropriately coarse grained delta function  is equivalent to the density of states, $|\rho(i)|^2$, leading to the final answer,
\begin{align}
	\left| \rho_0(j) \rho_0(k) \rho_0(i)\right|^2 \!\!\! \int\limits_{i_2,i_3}{\!\!\!}^{'} \left|\rho_0(i_2)\rho_0(i_3)\right|^2+ \left( \left| \rho_0(i)\rho_0(j)\rho_0(k)\right|^2 \!\! \int\limits_{i_1}{\!}^{'} \left|\rho_0(i_1)\right|^2\, \begin{Bmatrix}
        \mathcal O_k & \mathcal O_j & \mathcal O_i\\
        \mathcal O_k & \mathcal O_j & \mathcal O_1
    \end{Bmatrix} + S_3~{\rm perm.} \right)~.
\end{align}
There are additional diagrams that arise due to enhanced symmetry factors, coming from identifying different internal indices with the external indices. However, all such diagrams are further suppressed in $1/\rho_0 \sim e^{-c}$ and are therefore not studied here.

Thus, at leading order in the $e^{-c}$ expansion, the two-point function, or in other words, the second moment, of the $6j$-tensor model is given precisely by the expression in the Gaussian ensemble of \cite{Chandra:2022bqq}, while the four-point function receives additional contributions from the non-Gaussian terms we introduced in this work. Our constrained matrix/tensor model thus gives a non-Gaussian generalization of the ensemble, whose interpretation in terms of a simplicial action of the 6j symbol of Virasoro is very striking.

\section{Discussion}\label{sec.discussion}
In this paper we have introduced the concept of approximate CFTs, defining them as theories that correspond to a set of CFT data (OPE coefficients and spectrum) that approximately satisfy some restricted set of CFT constraints, as opposed to true CFTs which exactly satisfy the full set of constraints. We gave a detailed description of this concept in section \ref{sec.approxCFTs}. In particular, it is not possible to distinguish this data from the data corresponding to an exact CFT using only `light' observations, that is only involving correlation functions of `light' data. Indeed a cutoff between what constitutes `light' and `heavy' data in the spectrum is part and parcel of the definition of an approximate CFT. We note that in the context of holographic CFTs such a cutoff is particularly natural, and should scale linearly in the central charge, i.e. linearly in the number of CFT local degrees of freedom.

Taking a Wilsonian outlook, it is not possible to distinguish such a set of CFT data from an exact UV-complete CFT using the observations we have access to, and therefore a physical observer can stay agnostic about this difference. As argued in section \ref{sec.approxCFTs}, our definition is partly inspired by the bootstrap approach to CFTs, and partly by the effective field theory of the dual gravitational description when it exists (i.e. for the case of holographic CFTs). One might be worried that approximate CFT data, as per our definition, might not even correspond to a physical quantum theory at all, but in fact this is not disheartening. With the recent advancement of our understanding of the effective field theory of gravity in two dimensions being described as an ensemble of  boundary quantum theories, this provides a physically motivated `space of theories' in higher dimensions, which moreover are indistinguishable in the IR but are compatible with a description by exact CFT data in the UV. We remind the reader that one very explicit construction of such approximate CFTs indeed comes from exact CFTs, where we alter `by hand' a number of conformal operator dimensions away from their exact values. That this procedure indeed defines an approximate CFT according to our definition is shown in detail in section \ref{sec.example-approxCFT}, where we give an explicit example.

While the issue of the role of low-energy supergravity versus its UV completion constitutes one important motivation behind our work, it is far from the only one.  A second major field of application comes from the question of what becomes of random-matrix theory, which is so successful in describing chaotic many-body systems, when we move to quantum field theories and in particular those with conformal symmetry. This adds the simultaneous complications of locality and a high degree of symmetry to the problem, so the question of the definitions of the (elements of the) ensemble which is to be averaged becomes pertinent. In this context, the notion of an approximate CFT equips us with a space over which one can describe an ensemble for the purpose of understanding an effective description of quantum ergodic aspects of conformal field theories, i.e. their adapted notion of RMT universality. 

Conceptually it is important to keep in mind that in the case of physical quantum theories, i.e. those given given by a fixed set of CFT data, we may interpret both their quantum ergodic behavior as well as their semi-classical gravitational description (when it exists) as arising from  a coarse-graining of the relevant part of the spectrum.
This means that in spite of these theories being defined in the UV as individual non-averaged models, an ensemble description provides an efficient tool to model the physics of interest in such cases \cite{Altland:2020ccq}. This is not different from how an ensemble description efficiently models the spectrum of heavy nuclei or impurities in disordered systems. The question of whether such an averaging is fundamental or only an artifact of the effective description remains nevertheless extremely pertinent in the case of gravity.

Subsequently, we have described an ensemble of CFT data in section \ref{sec.var-cross} specifically for the purpose of studying gravity in three dimensions. This ensemble is directly defined in terms of the moments of the CFT data and is constrained by requiring that the CFT constraints be obeyed to increasingly higher accuracy. As explained previously, this ensemble differs in a subtle way from an ensemble over approximate CFTs in that it permits CFT data that violates `light' constraints by an arbitrary amount, a feature prohibited in approximate CFTs. However, in practice this is not an issue because such outliers are heavily suppressed in the ensemble. One may think of this as giving a {\it canonical} description where outliers are merely exponentially suppressed, as opposed to the previous {\it microcanonical} description where such outliers are strictly cut off.

Lastly, in section \ref{sec.tensor} we have defined a tensor model that gives an explicit ensemble of operator dimensions and CFT structure constants which approximately satisfies the constraints of a two-dimensional CFT. In particular this ensemble approximately satisfies modular crossing of torus one-point functions, as well as four-point crossing of local operators on the sphere. Interestingly the corresponding ensemble, when restricted to the operator content relevant for pure three-dimensional gravity -- that is the spectrum contains only the identify Viraosoro module and the continuum `black-hole' states above $(h,\bar h) = \frac{c-1}{24}$, gives rise to an action that strongly resembles that of three-dimensional simplicial gravity. To be precise, the potential of the OPE coeffcients implied by the average satisfaction of CFT constraints gives rise to a generalization of the `Boulatov' action of simplicial gravity, where the ${\cal U}_q(SU(2))$ Racah-Wigner coefficients are replaced by those of Virasoro, and three-simplices are labeled by Virasoro representations instead of ${\cal U}_q(SU(2))$. Moreover, we have an additional quartic vertex involving the pillow contraction structure of the tensor indices, which is likely to be important in preventing melonic dominance in the diagrammatic expansion. In addition to the tensor sector, our model also integrates over the matrix(-es) of  conformal dimensions.  This ensemble is similar to the one defined in section \ref{sec.var-cross} albeit for a finite spectrum. 

 In principle, one could have defined a matrix and tensor ensemble that included a smoothly varying potential in addition to the one resulting from the squares of the crossing constraints. This would also localize to solutions to the constraints in the scaling limit, but would not be an unbiased distribution. Therefore the  conjecture that our tensor model reproduces pure 3d gravity is equivalent to the claim that in this sense pure 3d gravity corresponds to the maximal ignorance  ensemble consistent with conformal symmetry and all crossing constraints.\footnote{See also \cite{DiUbaldo:2023qli} for a similar observation.} 
The model, as defined, gives only a discrete approximation of a given Euclidean three manifold, but we believe that an appropriate double scaling (or perhaps triple) can be defined in order to obtain a continuum description. One might then expect to find a gravitational description in the continuum limit of this simplicial gravity, which will be equivalent to an ensemble like the one we described in section \ref{sec.var-cross}. This exciting prospect is something we leave to be addressed in the future, together with an in-depth study of the tensor model of section \ref{sec.tensor} and its relation to three-dimensional gravity.

An important question is whether many diagrams of the tensor model must be summed, in a limit corresponding to dense triangulations, to obtain the contribution of a single 3-manifold topology as in simplicial gravity \cite{Boulatov:1992vp}, or whether a single Feynman graph can directly give the partition function with that topology, as in the formulation of 3d gravity as topological quantum field theory \cite{andersen2011tqft, andersen2013new, Collier:2023fwi}. Note that the tensor model amplitudes involving vertices with the 6j symbol of the Virasoro algebra and propagators given by the $C_0$ function are exactly of the form appearing in the Virasoro TQFT calculation \cite{Collier:2023fwi} of 3d gravity. One intriguing possibility is that in the triple scaling limit of the tensor model, in which the delta functions appearing in the potential become exact and the spectral range goes to infinity, the orthogonality relation and pentagon identity of the 6j symbols lead to an exact degeneracy between many Feynman diagrams, providing a connection between these two alternatives. Precisely such a phenomenon appeared in the matrix model for JT gravity with matter \cite{Jafferis:2022wez}.

We will now discuss some remaining aspects of the three approaches or usages of our concept of approximate CFT put forth in this paper. 
\subsection*{Does gravity ever violate crossing?}\label{sec.crossing.gravity}

In our definition of approximate CFTs, we have allowed for the possibility of violating crossing (or modular invariance) by bounded amounts. The question we would now like to discuss is whether such a violation is ever reproduced in a gravity calculation.
One could argue that any bulk calculation must always yield an answer that is manifestly crossing/modular invariant, and the true concern is with unitarity. This does however rely on some assumptions, as we will explain below.

At the level of local correlation functions, it is certainly true that perturbatively, we obtain crossing-invariant answers from gravity as Witten diagrams are computed by explicitly summing over channels, resulting in a crossing-invariant answer by design. Non-perturbatively, the situation is more subtle. Naively, asymptotic boundary conditions are given by the boundary metric plus the operator insertions. No prescription is ever given to pick a channel, and it would seem thus natural that crossing invariance prevails non-perturbatively. This question is related to correlation functions in Mellin space \cite{Gopakumar:2016wkt}, which are manifestly crossing-symmetric but not necessiraly unitary (see \cite{Penedones:2019tng} for a discussion of the existence of non-perturbative Mellin correlators). The argument becomes even cleaner when we consider modular invariance. The statement of modular invariance from the gravity point of view is the equivalence between two solid tori, which are mapped into one another by large diffeomorphisms. There, it is very clear that the two tori are treated exactly identically from the point of view of gravity if the gravity path integral is instructed to sum over all geometries with a given boundary condition. From this point of view, we wouldn't expect any violations of modular invariance, at genus one or at any higher genus. 

However, there are hidden assumptions in the statements made above. For example, this assumes that the correct nature of the gravitational path integral is to sum over large diffeomorphism that map equivalent tori (or operator insertions) into one another. It is logically possible that another condition must be input in the gravitational path integral. For example, one may need to pick a single bulk topology (or channel decomposition). There are different choices that can be made for which one to pick (which all yield the same answer), but one should not sum over them. This possiblity was advocated for in the tensionless string \cite{Eberhardt:2020bgq,Eberhardt:2021jvj}. In such a case, modular invariance is no longer automatic, but is ultimately restored by detailed information of the dynamics (see also \cite{Benini:2022hzx} for a similar scenario).

Making progress on this front remains an important goal. In this paper, we have taken a safer approach, and do not commit on whether or not crossing/modular symmetry is satisfied non-perturbatively, which is one of the roles of our tolerance parameter.

\subsection*{Non-Gaussianities are always there}

When studying the higher moments (in particular the variance) of the crossing equation, we have seen that non-Gaussiniaties of OPE coefficients play an important role. These non-Gaussiniaties were already known and discussed in detail in \cite{Belin:2021ryy}. The net weight of the non-Gaussianities are exponentially suppressed in the entropy, but they can build up to large and dominant effects (see also \cite{Foini:2018sdb,Dymarsky:2018ccu,Richter:2020bkf,Wang:2021mtp} in the context of eigenstate thermalization). This was known from \cite{Belin:2021ryy} in the context of higher genus partition functions. In this paper, we have shown that the same non-Gaussianities enter into the higher moments of the modular crossing equations on lower Riemann surfaces.

This is perhaps not too surprising, as both quantities involve higher moments of the OPE coefficients, but the structure is quite elegant. In fact, this structure plays a crucial role in the tensor model for 3d gravity. Two unique objects - the modular and four-point crossing kernels - encode everything and build up the dual bulk geometry by gluing elementary building blocks together.

\subsection*{Near threshold states and a statistical modular bootstrap?}

In section \ref{sec.var-cross}, we studied ensemble-averaged observables. As already pointed out in \cite{Chandra:2022bqq}, even in the Gaussian model, the ensemble-average of squares of conical defect four-point functions are invariant under simultaneous crossing transformations of the cross-ratios $x$ and $x'$. This follows from the crossing invariance of Liouville correlation functions, which implies that \rref{ss} and \rref{tt} are equal to each other under this simultaneous transformation. 

This property is highly non-trivial. To see this, consider the OPE limit of the $t$-channel: $1\to1$, $2\to2$ where the two cross-ratios $x$ and $x'$ are equal (i.e. $x=\bar{x}\to1$). In this channel, the high-energy states are highly suppressed by the kinematics, so one would naively expect them not to contribute much to the square of the four-point function. In the cross-channel, the high energy states are not suppressed and the cross-Wick contraction of the OPE coefficients of the $s$-channel $C_{12k}$ produce the Liouville correlation function. But the two channels are equal! This means something must be contributing in the $t$-channel as well.

There are essentially two possibilities of how this happens. The first possibility is that there is no saddle contribution in the sum over energies, which roughly means that the $s$-channel receives contributions from all energies with no peaked contribution. The second possibility is that there is a saddle-energy, which gets pushed closer and closer to the black hole threshold. 

This second possibility is very interesting from a modular bootstrap point of view \cite{Hellerman:2009bu}. Even if we are dealing with an ensemble here, we believe that many properties of this ensemble are also displayed in a single chaotic CFT, where one views the averaging as a statistical coarse-graining over the high-energy spectrum. Could one formulate a statistical modular bootstrap, and improve this way the current bounds (see \cite{Hartman:2019pcd}) all the way down to the black hole threshold? It would be interesting to study this further.

\subsection*{Connection between approximate CFTs and our tensor model}
We conclude this discussion by highlighting some aspects of the tensor model, and in particular its exact relationship to the concept of {\it approximate CFT} introduced in section \ref{sec.approxCFTs} of this paper. Broadly speaking the matrix/tensor model of section \ref{sec.tensor} is an explicit realization of a probability distribution over approximate CFTs, governed by a measure over matrix and tensor degrees of freedom. This measure is quartic in the tensors and fully specified by modular and four-point crossing constraints written in terms of the appropriate crossing kernels, together with a choice of spectrum. The resulting probability distribution is strongly peaked on configurations respecting the CFT constraints, but allows fluctuations around them, leading to non-trivial higher moments of the crossing equations. We note that strictly speaking there are tails in the measure which give an exponentially small, but nonzero probability density of violating the constraints by a large amount. Such excursions were ruled out in our definitions of section \ref{sec.approxCFTs}, but as alluded to above, one may view the difference as being akin to describing a thermal ensemble via a microcanonical approach (excursions are strictly cutoff by the microcanonical window), versus the canonical approach (excursions are strongly suppressed by the exponential measure). It may help to note that ordinary RMT also allow exponentially suppressed large excursions of eigenvalues, even though the underlying physical Hamiltonian is strictly bounded.

A second aspect that deserves comment is the fact that in our construction of section \ref{sec.tensor} we have both truncated the number of operators included in order to render the matrix and tensor degrees of freedom finite, and kept the coefficient $a$ `large but finite'. The idea is that an extrapolation to a continuum limit of the model involves a double scaling in which $a$ is taken to infinity, while taking the support of the spectral density (of operator dimensions) to infinity, holding an appropriate ratio finite. The details of this limit are beyond the scope of this article, but it is natural to expect an analysis along the lines of \cite{Jafferis:2022uhu,Jafferis:2022wez} to apply in this case as well. In the same spirit, it may be useful to point out that the analysis of section \ref{sec.var-cross} takes place in a `half double-scaled' limit, in the sense that there the number of participating operators is infinite, but the regulator $a$ is in its `large but finite' regime. Again, a similar half double-scaled limit was used at an intermediary stage of the analysis in \cite{Jafferis:2022uhu,Jafferis:2022wez}. We plan to return to these questions in future work.

Let us close on a technical remark, namely that the matrix/tensor model is built on `local crossing' equations, in the sense that the CFT constraints are formulated only in term of the crossing kernels, and not in terms of true observables like four-point functions (this was also advocated for in a different context in \cite{Collier:2019weq}). This turns out to be very convenient for the candidate model describing pure gravity where the crossing kernels involve integrals along a simple contour following the Liouville continuum. This makes the model very tractable and displays a strong resemblance to previous ideas on simplicial quantum gravity, but the precise connection certainly deserves further study. As we already mentioned, the regularized tensor model should view the local crossing equations as suitably smeared, which connects to the parameter $\zl$ of our definition of approximate CFTs, and can be viewed as a coarse-graining over the microstates. A full continuum limit most likely requires taking the smearing to zero at the same time as one performs the usual double-scaling, leading to a triple-scaling limit.  In the end, the fact that the model is ultimately based on the basic principles of the CFT bootstrap gives one hope that CFT techniques will be useful in approaching the continuum (gravity) limit. It would be interesting to study this further, as well as its generalizations to higher dimensions.

\acknowledgments
We thank Tom Hartman, David Kolchmeyer, Diego Liska, Juan Maldacena, Alex Maloney, Greg Mathys, Henry Maxfield, Marco Meineri, Baur Mukhametzhanov, Eric Perlmutter, Boris Post, Liza Rozenberg, Herman Verlinde, Edward Witten, Gabriel Wong, Sasha Zhiboedov. This work has been supported in part by the Fonds National Suisse de la Recherche Scientifique (Schweizerischer Nationalfonds zur F\"orderung der wissenschaftlichen Forschung) through Project Grant 200020182513, and the NCCR 51NF40-141869, The Mathematics of Physics (SwissMAP). JdB is supported by the European Research Council under the European Unions Seventh Framework Programme (FP7/2007-2013), ERC Grant agreement ADG 834878. The work of DLJ is supported in part by the DOE grant DE-SC0007870.

\appendix

\section{A simple toy model}\label{app.ToyModel}
In this appendix we describe briefly the relation between the bare propagator and full two-point function for the example of a vectorized ordinary integral, as used in the illustrative toy model of section \ref{sec.tensor}. Let us consider the theory
\begin{equation}\label{eq.action1}
    Z[J] = \int \left[ d\phi \right] e^{-S_\phi + J \cdot \phi}\,,
\end{equation}
where $\phi_i$ is an $N$-component vector and the action $S_\phi$ is the O$(N)$ invariant expression
\begin{equation}\label{eq.action2}
    S_\phi = \frac{1}{2}\phi^T K^{-1}\phi + \frac{\lambda}{4 N} \left( \phi^T \phi \right)^2\,,
\end{equation}
with $K_{ij} = K \delta_{ij}$ is the bare propagator. The measure is given by the product of individual measures $[d\phi] = \prod_i d\phi_i$.
\subsection{Schwinger-Dyson equations and exact two-point function}
We now derive the full hierarchy of Schwinger Dyson equations for un-normalized correlation functions of the type
\begin{equation}
    G^{(2n)}_{i_1 i_2\,\ldots i_{2n}} = \left\langle \phi_{i_1} \phi_{i_2} \cdots \phi_{i_{2n}} \right \rangle\,,
\end{equation}
where we have of course in mind to eventually consider only O$(N)$-invariant cases by summing over free indices in pairs. To arrive at the set of SD equations for this toy model, we consider the identity
\begin{equation}
    0 = \int [d\phi] \frac{d}{d\phi_i} e^{-S_\phi + J \cdot \phi} = \left(-\frac{\delta S_\phi}{\delta \phi_i} \Bigr|_{\phi_i \rightarrow \partial_{J_i}}  + J_i \right) Z[J]
\end{equation}
Here we have simply assumed that the boundary terms of the first expression vanish, and subsequently used the fact that in the presence of sources we may replace each occurrence of $\phi_i$ by a derivative with respect to the appropriate source $\frac{\partial}{\partial J_i}$. This expression can then entirely be moved outside the integral, which reduces to the generating functional itself, as shown. In order to get SD equations for any-order correlation functions one now simply takes derivatives with respect to sources as needed. For the quartic toy model at hand, we obtain the SD equations
\begin{equation}
    G^{(2n)}_{i_1 \ldots i_{2n}} = -\frac{\lambda}{N}K G^{(2n+2)}_{i_1 ii \ldots i_{2n}} + \sum_k K \delta_{i_1 i_k} G^{(2n-2)}_{\{i_2\ldots i_{2n}\} \backslash i_k}\,,
\end{equation}
where the notation, ${\{i_2\ldots i_{2n}\} \backslash i_k}$, of the final term indicates that we take the set of indices $\{i_2, i_3 , \ldots i_{2n}\}$ and exclude the single index $i_k$ which is featured on the Kronecker delta preceding the correlator. Specializing to the O$(N)$-invariant two-point function we thus have
\begin{eqnarray}
    G^{(2)}_{ii} = -\frac{\lambda}{ N}G^{(4)}_{ii jj} + N K G^{(0)}\,.
\end{eqnarray}
In this expression we sum over all repeated indices and we may identify the zero-point function as the partition function that allows us to write the SD for normalized correlation functions as
\begin{equation}\label{eq.twoPtSD}
     \hat G^{(2)}_{ii} = - \frac{\lambda}{ N}\hat G^{(4)}_{ii jj} + N K \,.
\end{equation}
We may now proceed to solve this equation at large $N$, where we shall see it closes on the normalized two-point function. Firstly let us consider the $N$ counting of the various terms. The left-hand side of \eqref{eq.twoPtSD} is ${\cal O}(N)$ due to the single sum over indices. The first term on the right-hand side equally is of ${\cal O}(N)$ due to the two sums being compensated by the single $N^{-1}$ from the coupling. Finally the bare propagator term explicitly is ${\cal O}(N)$. Thus, the two-point SD equation finally takes the diagrammatic form
\begin{equation}
     \includegraphics[width=0.7\textwidth]{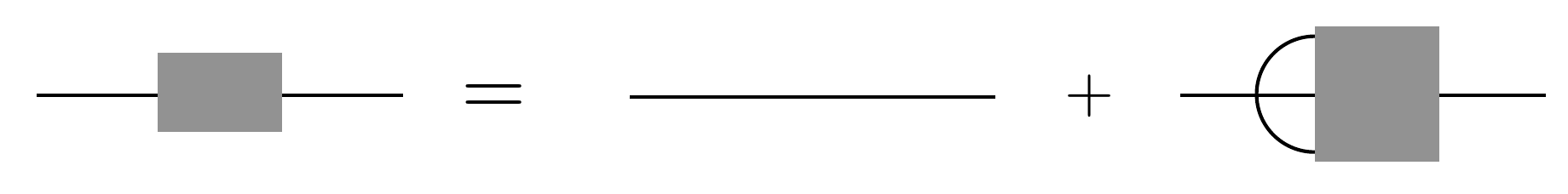}\,.
\end{equation}
The left-hand side is the full two-point function, and the right-hand side includes the bare propagator contribution, as well as the full four-point function. Note the formal similarity with the tensor model SD equation \eqref{eq.SDsimple}, shown diagrammatically in Figure \ref{fig.TensorSDEquation} of the main text. 
Using now the large-$N$ factorization property
\begin{eqnarray}
    \hat G^{(4)}_{ii jj} =  \hat G^{(2)}_{ii}  \hat G^{(2)}_{jj} + {\cal O} \left( 1 /N \right)\,,
\end{eqnarray}
we can close the SD equation for the two-point function
\begin{equation}\label{eq.largeNSDtoy}
    \hat G^{(2)}_{ii} = -\frac{\lambda}{ N} K \left( G^{(2)}_{ii}  \right)^2 + N K \qquad \Rightarrow \qquad \hat G^{(2)} = \frac{1}{2K \lambda} \left(1\mp \sqrt{1+4 K^2\lambda}  \right) = \left\{ \begin{array}{c}
          K + {\cal O} \left( \lambda \right) \\
          -\frac{1}{K \lambda}+ {\cal O} \left( 1 \right)
    \end{array} \right.  \,,
\end{equation}
where we have defined the $O(N)$-invariant two-point function
\begin{equation}
    \hat G^{(2)} := \frac{1}{N} \sum_i \hat G_{ii}^{(2)}\,.
\end{equation}
For the case of interest, namely where the model \eqref{eq.action1} arises from exponentiating a quadratic constraint, the first of these solutions corresponds to the `bare propagator', while the second one is the full resummed two-point function and always coincides with the solution of the quadratic constraint itself.

The results of this calculation thus ensure that whichever form of the potential we choose to impose the quadratic constraint (see section \ref{sec.fullbare} of the main part of this paper), the full two-point function is always given by the solution of the quadratic constraint itself.

\bibliographystyle{ytphys}
\bibliography{ref}

\end{document}